\documentclass[twocolumn,10pt,oneside,reqno]{article}

\usepackage[top=1in, bottom=1in, left=.65in, right=.65in]{geometry}
\usepackage{amsmath, amsthm, amssymb}
\usepackage{graphicx}
\usepackage{epstopdf}
\usepackage{hyperref}
\usepackage{overpic}
\usepackage{cancel}
\usepackage{rotating}
\usepackage{url}
\usepackage{caption}
\usepackage{color}
\usepackage[usenames,dvipsnames,svgnames,table]{xcolor}
\usepackage{rotating}
\usepackage{multirow}
\usepackage{wrapfig}
\usepackage{multicol}
\usepackage{multirow}
\usepackage{adjustbox}
\usepackage{setspace}
\usepackage{palatino} % needs 10
\usepackage{algorithm,algcompatible}
\usepackage{algpseudocode}
\usepackage{tikz}
\usepackage{booktabs}
\renewcommand{\Comment}[2][.4\linewidth]{%
	\leavevmode\hfill\makebox[#1][l]{$\triangleright$~#2}}
\usepackage{mathtools}  
\usepackage{psfrag}
\usepackage{tikz}
\usepackage[T1]{fontenc}
\usepackage{array}
\usepackage{makecell}
\newcolumntype{x}[1]{>{\centering\arraybackslash}p{#1}}

\newcommand\diag[4]{%
	\multicolumn{1}{p{#2}|}{\hskip-\tabcolsep
		$\vcenter{\begin{tikzpicture}[baseline=0,anchor=south west,inner sep=#1]
			\path[use as bounding box] (0,0) rectangle (#2+2\tabcolsep,\baselineskip);
			\node[minimum width={#2+2\tabcolsep},minimum height=\baselineskip+\extrarowheight] (box) {};
			\draw (box.north west) -- (box.south east);
			\node[anchor=south west] at (box.south west) {#3};
			\node[anchor=north east] at (box.north east) {#4};
			\end{tikzpicture}}$\hskip-\tabcolsep}}
\usepackage[bottom,flushmargin,hang,multiple]{footmisc}
\usepackage{lipsum}

\algnewcommand\algorithmicinput{\textbf{Input:}}
\algnewcommand\INPUT{\item[\algorithmicinput]}
\algnewcommand\algorithmicoutput{\textbf{Output:}}
\algnewcommand\OUTPUT{\item[\algorithmicoutput]}
\algnewcommand\algorithmicoptional{\textbf{Optional:}}
\algnewcommand\OPTIONAL{\item[\algorithmicoptional]}

\newcommand{\bA}{\mathbf{A}}
\newcommand{\bAY}{\mathbf{A_Y}}
\newcommand{\bAtilde}{\mathbf{\tilde{A}}}
\newcommand{\bAtildeY}{\mathbf{\tilde{A}_Y}}
\newcommand{\bB}{\mathbf{B}}
\newcommand{\bBhat}{\mathbf{\hat{B}}}
\newcommand{\bBY}{\mathbf{B}_\bY}
\newcommand{\bBYhat}{\bBhat_\bY}
\newcommand{\bBtilde}{\mathbf{\tilde{B}}}

\newcommand{\bb}{\mathbf{b}}
\newcommand{\bC}{\mathbf{C}}
\newcommand{\bF}{\mathbf{F}}
\newcommand{\bG}{\mathbf{G}}
\newcommand{\bP}{\mathbf{P}}

\newcommand{\bq}{\mathbf{q}}
\newcommand{\rtilde}{\tilde{r}}
\newcommand{\bR}{\mathbb{R}}
\newcommand{\bS}{\mathbf{S}}
\newcommand{\bs}{\mathbf{s}}

\newcommand{\bU}{\mathbf{U}}
\newcommand{\bUY}{\mathbf{U_Y}}
\newcommand{\bUhat}{\mathbf{\hat{U}}}
\newcommand{\bUhatY}{\mathbf{\hat{U}_Y}}
\newcommand{\bUtilde}{\mathbf{\tilde{U}}}
\newcommand{\bUtildeX}{\mathbf{\tilde{U}}}
\newcommand{\bUtildeY}{\mathbf{\tilde{U}_Y}}
\newcommand{\bu}{\mathbf{u}}
\newcommand{\bV}{\mathbf{V}}
\newcommand{\bVY}{\mathbf{V_Y}}
\newcommand{\bVhat}{\mathbf{\hat{V}}}
\newcommand{\bVhatY}{\mathbf{\hat{V}_Y}}
\newcommand{\bVtilde}{\mathbf{\tilde{V}}}
\newcommand{\bVtildeY}{\mathbf{\tilde{V}_Y}}
\newcommand{\bW}{\mathbf{W}}
\newcommand{\bw}{\mathbf{w}}
\newcommand{\bWY}{\mathbf{W_Y}}

\newcommand{\bX}{\mathbf{X}}
\newcommand{\bx}{\mathbf{x}}

\newcommand{\bxtilde}{\mathbf{\tilde{x}}}
\newcommand{\bY}{\mathbf{Y}}
\newcommand{\by}{\mathbf{y}}

\newcommand{\bLambda}{\boldsymbol{\Lambda}}
\newcommand{\bLambdaY}{\boldsymbol{\Lambda}_\bY}
\newcommand{\blambda}{{\lambda}}

\newcommand{\bnu}{\nu}
\newcommand{\bOmega}{\boldsymbol{\Omega}}
\newcommand{\bOmegaY}{\boldsymbol{\Omega_\bY}}
\newcommand{\bPhi}{\boldsymbol{\Phi}}
\newcommand{\bPhiY}{\boldsymbol{\Phi_\bY}}
\newcommand{\bPhiS}{\boldsymbol{\Phi_\bS}}
\newcommand{\bphi}{\boldsymbol{\phi}}
\newcommand{\bphiS}{\boldsymbol{\phi_\bS}}
\newcommand{\bphiY}{\boldsymbol{\phi_\bY}}

\newcommand{\bPsi}{\boldsymbol{\Psi}}
\newcommand{\bSigma}{\boldsymbol{\Sigma}}
\newcommand{\bSigmaY}{\boldsymbol{\Sigma_\bY}}
\newcommand{\bSigmahat}{\boldsymbol{\hat{\Sigma}}}
\newcommand{\bSigmahatY}{\boldsymbol{\hat{\Sigma}_\bY}}
\newcommand{\bSigmatilde}{\boldsymbol{\tilde{\Sigma}}}
\newcommand{\bSigmatildeY}{\boldsymbol{\tilde{\Sigma}_\bY}}
\newcommand{\bTheta}{\boldsymbol{\Theta}}
\newcommand{\bUpsilon}{\boldsymbol{\Upsilon}}
\newcommand{\bxi}{\xi}

\definecolor{blue}{rgb}{0,0,1}
\definecolor{darkgreen}{rgb}{0,0.5,0}
\definecolor{red}{rgb}{1,0,0}
\definecolor{teal}{rgb}{0,0.5,0.7}

\newtheorem{theorem}{Theorem}
\newtheorem{lemma}{Lemma}
\newtheorem{corollary}{Corollary}
\newtheorem{assumption}{Assumption}

\graphicspath{{Figures/}}
\newcommand{\boundellipse}[3]% center, xdim, ydim
{(#1) ellipse (#2 and #3)
}
\title{\huge{Dynamic mode decomposition\\ for compressive system identification}
}
\author{Zhe Bai$^{1*}$, Eurika Kaiser$^{1}$, Joshua L. Proctor$^{3}$, J. Nathan Kutz$^{2}$, Steven L. Brunton$^{1}$\\~\\
\small{$^1$ Department of Mechanical Engineering, University of Washington, Seattle, WA 98195, United States}\\
\small{$^2$ Department of Applied Mathematics, University of Washington, Seattle, WA 98195,
United States}\\
\small{$^3$ Institute of Disease Modeling, Bellevue, WA 98004, United States}
\\~\\
}
\date{}
\date{\today}
\begin{document}

\twocolumn[
  \begin{@twocolumnfalse}

\maketitle

%\blfootnote{$^*$ Corresponding author. Tel.: +1 609 921 6415.\\ {\indent\emph{E-mail address:} sbrunton@uw.edu (S.L. Brunton).}}
%%%%%%%%%%%%
%%% ABSTRACT
%%%%%%%%%%%%
\begin{abstract}
Dynamic mode decomposition has emerged as a leading technique to identify spatiotemporal coherent structures from high-dimensional data, benefiting from a strong connection to nonlinear dynamical systems via the Koopman operator.  
In this work, we integrate and unify two recent innovations that extend DMD to systems with actuation~\cite{Proctor2016siads} and systems with heavily subsampled measurements~\cite{Brunton2015jcd}.  
When combined, these methods yield a novel framework for compressive system identification~\footnotemark.  
It is possible to identify a low-order model from limited input--output data and reconstruct the associated full-state dynamic modes with compressed sensing, adding interpretability to the state of the reduced-order model.  
Moreover, when full-state data is available, it is possible to dramatically accelerate downstream computations by first compressing the data.  
We demonstrate this unified framework on two model systems, investigating the effects of sensor noise, different types of measurements (e.g., point sensors, Gaussian random projections, etc.), compression ratios, and different choices of actuation (e.g., localized, broadband, etc.).  
In the first example, we explore this architecture on a test system with known low-rank dynamics and an artificially inflated state dimension.  
The second example consists of a real-world engineering application given by the fluid flow past a pitching airfoil at low Reynolds number.  
This example provides a challenging and realistic test-case for the proposed method, and results demonstrate that the dominant coherent structures are well characterized despite actuation and heavily subsampled data. \\

\end{abstract}
~\\
  \end{@twocolumnfalse}
 ]
\footnotetext{Code is publicly available at: \url{https://github.com/zhbai/cDMDc}}

%%%%%%%%%%%%%%%%%%%%
%% NOMENCLATURE
%%%%%%%%%%%%%%%%%%%%

%%%%%%%%%%%%%%%%%%%%
%% NOMENCLATURE
%%%%%%%%%%%%%%%%%%%%
\begin{figure*}
\framebox[\textwidth]{
\noindent\begin{minipage}[h]{.45\textwidth}
\section*{Nomenclature}
\begin{tabular}{ll}
$\bA, \bAtilde$ & State transition matrix for $\bx$ and $\tilde{\bx}$ \\
$\bAY, \bAtildeY$ & State transition matrix for $\by$ \\
$\bB, \bBtilde, \bBhat$ & Actuation matrix\\ %($\bBhat$ is estimate)\\
$\hat{\bB}$ & Estimate of $\bB$\\ %($\bBhat$ is estimate)\\
%$\bBY, \bBtildeY, \bBYhat$ & Actuation matrix in $\by$\\
%$\bBhat, \bBYhat$ & Estimated control matrix of $\bX$, $\bY$ \\
$\bb$ & Vector of DMD amplitudes \\
$\bC$ & Output measurement matrix \\
$\bF$ & Discrete cosine transform (DCT)\\
%$\bG$ & Augmented matrix containing $\bA$ and $\bB$ \\
$\bG$ & Augmented matrix of $\bA$ and $\bB$\\% $[\bx; \bu]$ \\
$K$ & Number of nonzero DCT coeffs. \\
%$i$ & Index of individual \\
%$l$ & Dimension of the input variable \\
%$m$ & Number of data snapshots \\
%$n$ & Dimension of the state, $\bx \in \bR^n$ \\
$\bP$ & Projection matrix acting on $\bx$ \\
$\bq$ & Spatial coordinate\\
%$p$ & Dimension of the measurement or output variable, $\by \in \bR^p$ \\
$r,\rtilde$ & Rank of truncated SVD of $\bX, \bOmega$\\
$\bs$ & DCT coefficients of $\bx$\\
$t$ & Time \\
$\Delta t$ & Time step \\
$t_k$ & $k$th discrete time step \\
$\bU,\bUhat,\bUtilde$ & Left singular vectors (POD modes)\\
& of $\bX, \bX', \bOmega$ \\
$\bUY,\bUhatY,\bUtildeY$ & Left singular vectors (POD modes)\\
& of $\bY, \bY', \bOmegaY$ \\
$\bu$ & Control input, $\bu \in \bR^q$ \\
$\bV,\bVhat,\bVtilde$ & Right singular vectors of $\bX, \bX', \bOmega$ \\
$\bVY,\bVhatY,\bVtildeY$ & Right singular vectors of $\bY, \bY', \bOmegaY$ \\
$\bW, \bWY$ & Eigenvectors of $\bAtilde,\bAtildeY$ 
\end{tabular}
\end{minipage}
\begin{minipage}[h]{.545\textwidth}
\begin{tabular}{ll}
%$\bw$ & Eigenvector of $\bAtilde$ \\
$\bX$ & Data matrix of states, $\bX \in \bR^{n\times m}$\\
$\bX'$ & Shifted data matrix, $\bX' \in \bR^{n\times m}$ \\
$\tilde{\bX}$ & Data matrix of reduced states, $\tilde{\bX} \in \bR^{r\times m}$\\
$\bxi$ & Spatial variable \\
$\bx$ & State vector, $\bx \in \bR^{n}$ \\
$\tilde{\bx}$ & Low-rank state vector, $\tilde{\bx} \in \bR^{r}$\\%Reduced-order vector of states
$\bY$ & Data matrix of measurements, $\bY \in \bR^{p\times m}$ \\
$\bY'$ & Shifted matrix of measurements, $\bY' \in \bR^{p\times m}$ \\
$\by$ & Output vector, $\by \in \bR^p$ \\
%\end{tabular}
%\end{minipage}
%\begin{minipage}[h]{.5\columnwidth}
%\begin{tabular}{ll}
$\eta$ & Noise magnitude \\
$\blambda$ & DMD eigenvalue of $\bA$\\
$\bnu$ & Wavenumber \\
$\bLambda,\bLambdaY$ & Matrix of DMD eigenvalues of $\bA, \bAY$\\
$\bOmega$ & State and input snapshot matrices $[\bX^T  \enspace  \bUpsilon^T]^T$\\% \in \bR^{(l+n)\times(m-1)}$ \\
$\bOmegaY$ & Output and input snapshot matrices $[\bY^T \enspace  \bUpsilon^T]^T$\\
%&$[\bY^T \enspace  \bUpsilon^T]^T$\\ %\in \bR^{(l+p)\times(m-1)}$ \\
$\omega$ & Continuous-time DMD eigenvalue of $\bA$\\
%s, $\omega \triangleq {\log(\lambda)}/{\Delta t}$ \\
$\bPhi,\bPhiY$ & Matrix of DMD (DMDc) modes of $\bA, \bAY$\\
$\bPhiS$ & Matrix of sparse representation of $\bPhi$ \\
$\hat{\bPhi}$ & Compressed sensing estimate of $\bPhi$\\
$\bPsi$ & Orthonormal basis (e.g. Fourier or POD) \\
$\bSigma,\bSigmahat,\bSigmatilde$ & Matrix of singular values of $\bX,\bX',\bOmega$ \\
$\bSigmaY,\bSigmahatY,\bSigmatildeY$ & Matrix of singular values of $\bY,\bY',\bOmegaY$ \\
$\bTheta$ & Product of measurement matrix and\\
&  sparsifying basis, $\bTheta=\bC\bPsi$ \\
$\bUpsilon$ & Data matrix of control inputs
\end{tabular}
\end{minipage}
}
\vspace{-0.2in}
\end{figure*}
\fontsize{10}{12}\selectfont
%%%%%%%%%%%%
%%% INTRODUCTION
%%%%%%%%%%%%
\section{Introduction}
Dynamic mode decomposition (DMD) is a dimensionality reduction technique introduced by Schmid~\cite{Schmid2010jfm} in the fluid dynamics community to decompose high-dimensional fluid data into dominant spatiotemporal coherent structures~\cite{Schmid2010jfm,Rowley2009jfm,Chen2012jns,Tu2014jcd,Kutz2016book,Brunton2017natcomm,Arbabi2016arxiv}.  
Shortly after its introduction, DMD was reframed by Rowley et al.~\cite{Rowley2009jfm} as a numerical technique to approximate the Koopman operator~\cite{Koopman1931pnas,Mezic2005nd,Mezic2013arfm,Bagheri2013jfm}, establishing a strong connection to the analysis of nonlinear dynamical systems.  
Unlike other dimensionality reduction techniques, such as proper orthogonal decomposition (POD)~\cite{HLBR_turb}, DMD is designed to extract modes that are spatially coherent, oscillate at a fixed frequency, and grow or decay at a fixed rate.  
Thus, DMD yields a set of modes along with a linear evolution model.  
Since being introduced in fluid dynamics, DMD has been widely applied in fields as diverse as epidemiology~\cite{Proctor2015ih}, neuroscience~\cite{brunton2016extracting}, robotics~\cite{Berger2014ieee}, video processing~\cite{Grosek2014arxiv}, and financial trading~\cite{Mann2016qf}.  

DMD may be thought of as a form of system identification, resulting in reduced-order models that are more tractable than the original high-dimensional dynamics.  
Low-order models are especially important for the control of high-dimensional dynamical systems, such as fluid flow control~\cite{Kaiser2014jfm,Brunton2015amr}, where fast control decisions must be enacted to reduce latency for robust performance. 
There are a number of excellent overviews of model reduction~\cite{Benner2015siamreview} and system identification~\cite{ljung1999SI,Leung1998ieee} for such high-dimensional systems.  
Balanced truncation provides a principled approach to reducing the system dimension by identifying a reduced-order model with the most jointly controllable and observable states~\cite{Moore1981ieeetac}, and extensions include the balanced proper orthogonal decomposition (BPOD) for systems with very large state dimension~\cite{Willcox2002aiaaj,Rowley2005ijbc}.  
It was recently shown~\cite{Ma2009tcfd} that BPOD is equivalent to the eigensystem realization algorithm (ERA)~\cite{Juang1985jgcd}.  
In Tu et al.~\cite{Tu2014jcd}, it was further shown that DMD may be equivalent to ERA under certain conditions.  
Proctor et al.~\cite{Proctor2016siads}, further extended DMD to include inputs and control, disambiguating internal state dynamics from the effect of actuation.  

DMD relies on the fact that even high-dimensional dynamics typically evolve on a low-dimensional attractor.  
This low-dimensional behavior suggests \emph{sparsity} in an appropriate basis, so that sparsity-promoting and randomized techniques may be exploited to reduce measurement resolution, bandwidth requirements, and computational overhead.  
Sparsity was first used in DMD by Jovanovi\'c et al.~\cite{Jovanovic2014pof} to select the dominant DMD modes.  
Compressed sensing~\cite{Candes2006ieee,Donoho2006ieee,Baraniuk2007ieee} was subsequently used to compute DMD using snapshots that were sampled below the Shannon-Nyquist sampling limit in time~\cite{Tu2014ef} and in space~\cite{Brunton2015jcd,Gueniat2015pof}.  
It was shown in Brunton et al.~\cite{Brunton2015jcd} that it is possible to reconstruct accurate DMD modes with surprisingly few spatial measurements, and if full-state data is available, performing DMD on compressed data dramatically reduces computation time.  
In addition to compressed sensing, randomized linear algebra has been leveraged to accelerate DMD computations~\cite{Erichson2017randomized,Bistrian2017randomized}.  

In this work, we combine the compressed sensing DMD~\cite{Brunton2015jcd} and DMD with control~\cite{Proctor2016siads} approaches, resulting in a powerful mathematical framework for compressive system identification.  
There has been other work on compressive system identification~\cite{Heckel2013arxiv,Kopsinis2010arxiv,Chen2009ieee,Gu2009ieee,Kalouptsidis2011sp,Naik2012conf,Bai2014aiaa,Brunton2016pnas,Kaiser2016arxiv,Kramer2015arxiv}, but this is the first effort in the context of DMD.  
We begin by presenting the DMD algorithm, and related compressed sensing and control extensions, in a common framework and notation, making it possible to unify these approaches.  
Combining these algorithms results in the compressive DMD with control (cDMDc) architecture.  
cDMDc then makes it possible to identify reduced-order models on downsampled spatial measurements of a high-dimensional system, and then reconstruct the full-state dynamic modes associated with the model using compressed sensing.  
This approach adds interpretability to otherwise black-box system identification models.  
The resulting cDMDc architecture is demonstrated on two example dynamical systems, including a high-dimensional fluid flow simulation.

%%%%%%%%%%%%
%%% BACKGROUND
%%%%%%%%%%%%
\begin{algorithm*}[t]
	\caption{Exact DMD~\cite{Tu2014jcd}} \label{alg DMD}
	\begin{algorithmic}[1]
		\INPUT{Data matrix $\bX$, shifted data matrix $\bX'$, and target rank $r$}.
		\OUTPUT{DMD spectrum $\bLambda$ and modes $\bPhi$}.  %\Comment Return DMD spectrum and modes.
		\Procedure{DMD}{$\bX,\bX', r$}
		\State $[\bU, \bSigma, \bV] \leftarrow$ SVD($\bX,r$) \Comment{Truncated $r$-rank SVD of $\bX$.}
		\State $\bAtilde \leftarrow \bU^*\bX'\bV\bSigma^{-1}$ \Comment{Low-rank approximation of $\bA$.}
		\State $[\bW, \bLambda] \leftarrow \text{EIG}(\bAtilde)$ \Comment{Eigen-decomposition of $\bAtilde$.}
		\State
		$\bPhi \leftarrow \bX'\bV\bSigma^{-1}\bW$	 \Comment{DMD modes of $\bA$.} 
		\EndProcedure
	\end{algorithmic}
 Note: If $\blambda_i=0$, then $\boldsymbol{\phi}_i = \bU\bw_i$ { for step } $5$.  In the original DMD algorithm~\cite{Schmid:2009} all modes are computed as $\boldsymbol{\phi}_i = \bU\bw_i$.  
\end{algorithm*}

\section{Background:  The Dynamic Mode Decomposition}\label{Sec:Background}

Matrix decomposition techniques are ubiquitous in the data sciences.  Their fundamental objective is often to extract low-rank and interpretable patterns from data.  Foremost among matrix decomposition methods is the {\em singular value decomposition} (SVD), which is the computational engine for {\em principal component analysis} (PCA) and produces a set of ranked orthonormal modes that capture the dominant correlation structures in the data.  However, the SVD fails to correlate both spatial and temporal features of the data together.  The DMD provides a least-square regression architecture whereby both space and time are jointly correlated by merging a spatial SVD with a temporal Fourier transform~\cite{Chen2012jns}.  Specifically, the DMD algorithm decomposes the data matrix ${\bf X}$ into the rank $r$ approximation
\begin{align}\label{eqn:b}
\bX \approx \bPhi \bLambda \bb
\end{align}
where the columns of $\bPhi$ are the DMD modes (spatial structures), the elements of the diagonal matrix $\bLambda$ are the corresponding DMD eigenvalues (with angular frequency $\lambda_i = \exp(\omega_i \Delta t)$), and the vector $\bb$ determines the weighting of each of the $r$ modes.   Thus, each spatial mode in $\bPhi$ is associated with a single temporal eigenvalue of $\bLambda$.

The DMD method was originally used as a low-rank diagnostic tool for decomposing fluid flow data into dominant spatiotemporal modes~\cite{Schmid2010jfm, Rowley2009jfm, Tu2014jcd, Kutz2016book}, providing a valuable interpretation of coherent structures in complex systems.   Recent innovations have also made significant progress to employ DMD for robust future-state prediction~\cite{Jovanovic2014pof,Askham2017arxiv} and control for input-output systems~\cite{Proctor2016siads}, even when using only a small number of measurements~\cite{Brunton2015jcd}.

The exact DMD algorithm formulates the decomposition as a least-squares regression.  
Specifically, a series of snapshots $\bx_k \in \bR^{n}$ sampled at discrete instances in time $t_k$, $k=0,\ldots,m$ are arranged into the data matrix
%
%\begin{subequations}
\begin{align} \label{Eqn:SnapshotMatrices}
\bX&=
\begin{bmatrix}
\vline & \vline & &\vline \\
\bx_0&\bx_1 &\hdots&\bx_{m-1}\\
\vline & \vline & &\vline 
\end{bmatrix}
\end{align}
and the time-shifted matrix 
\begin{align}
\bX'&=
\begin{bmatrix}
\vline & \vline & &\vline \\
\bx_1&\bx_2 &\hdots&\bx_m\\
\vline & \vline & &\vline 
\end{bmatrix}.
\end{align} 
%\end{subequations}
Typically $n\gg m$, i.e.\ there are many more spatial measurements available than temporal. 
The vector $\bx_k$ is the state of a high-dimensional system such as a fluid flow. 
Here, we assume evenly sampled snapshots, although this is generally not required~\cite{Tu2014jcd,Askham2017arxiv}.

The DMD algorithm constructs the leading eigendecomposition of the best-fit operator $\bA$, chosen to minimize $\| \bx_{k+1}-\bA\bx_k \|_2$ over the ${k=0,2,3,\cdots,{m-1}}$ snapshots, so that $\bX' \approx \bA \bX$.
Computationally, the matrix $\bA$ is obtained as
\begin{align} \label{Eqn:BestFitModel}
\bA = \bX' \bX^{\dagger}
\end{align}
where $\bA \in \bR^{n \times n}$ and $\bX^{\dagger}$ is the Moore-Penrose pseudo-inverse of $\bX$. 
The dominant eigenvectors of $\bA$ are the dynamic modes $\bPhi$, and the associated eigenvalues determine how these modes behave in time.
 
In practice, the high-dimensional $\bA$ is not computed directly.  Instead, the SVD can be used to first project to a low-rank subspace and then compute the matrix $\bAtilde = \bU^*\bA\bU$ which has many of the same eigenvalues as $\bA$.
This provides an efficient algorithm whose computational expense is bounded by the rank $r$ of the data.  The exact DMD algorithm of Tu {\em et al.}~\cite{Tu2014jcd} is shown in Algorithm~\ref{alg DMD}.   Using a variable projection optimization scheme, the exact DMD method can be modified to handle arbitrary temporal spacing between snapshots.  It further produces a more robust, or {\em optimal DMD} approximation, for noisy data~\cite{Askham2017arxiv}.

\begin{algorithm*}[t]
	\caption{DMD with control~\cite{Proctor2016siads}} \label{alg DMDc}
	\begin{algorithmic}[1]
		\INPUT{Data matrices $\bX$, $\bX'$, input snapshot matrix $\bUpsilon$, target rank $r$ of $\bX$ or $\bX'$ and $\rtilde$ of $\bOmega$.}
		\OPTIONAL{Actuation matrix $\bB$.}
		\OUTPUT{Spectrum $\bLambda$, modes $\bPhi$, [ and actuation matrix $\bBhat$].} \Comment{Optional outputs in brackets [$\cdot$].}
		\Procedure{DMDc}{$\bX,\bX',\bUpsilon,r,\rtilde,[\bB]$}
		\If{$\bB$ is known}
	     	\State $[\bLambda, \bPhi] \leftarrow$ DMD($\bX,\bX'-\bB\bUpsilon,r$) \Comment{Perform \emph{DMD} (Algorithm \ref{alg DMD}) adjusted}
	     	\State \Comment{for known actuation.}
	    	\Else
%		\State $\bOmega\in \bR^{(l+n)\times(m-1)} \leftarrow \begin{bmatrix}\bX \\ \bUpsilon \end{bmatrix}$ \Comment{Augmented matrix of the state and}
		\State $\bOmega \leftarrow \begin{bmatrix}\bX \\ \bUpsilon \end{bmatrix}$ \Comment{Matrix of the state and input snapshots.}	
		\State $[\bUtilde, \bSigmatilde, \bVtilde] \leftarrow$ SVD($\bOmega,\rtilde$) \Comment{Truncated $\tilde{r}$-rank SVD of $\bOmega$.}
%		\State $\bUtilde_{1} \in \bR^{n \times(m-1)}, \bUtilde_{2} \in \bR^{l \times(m-1)} \leftarrow \bUtildeX=\begin{bmatrix}
%			\bUtilde_{1}\\ \bUtilde_{2} \end{bmatrix}$  \Comment{Split $\bUtilde$ into two components.}
		\State $\bUtilde_{1},\; \bUtilde_{2}  \leftarrow \bUtildeX$  \Comment{Split $\bUtilde$ into two components.}
		\State $[\bUhat,\bSigmahat,\bVhat] \leftarrow$ SVD($\bX',r$) \Comment{Truncated $r$-rank SVD of $\bX'$.}
		\State $\bAtilde \leftarrow \bUhat^*\bX'\bVtilde\bSigmatilde^{-1}\bUtilde^*_{1}\bUhat$. \Comment{Low-rank approximation of $\bA$.}
%    \Comment{Approximation of $\bA$.}
		\State $\bBtilde \leftarrow \bUhat^{*}\bX'\bVtilde\bSigmatilde^{-1}\bUtilde^*_{2}$ \Comment{Estimate reduced actuation matrix $\bBtilde$.}
		\State $\bBhat \leftarrow \bX'\bVtilde\bSigmatilde^{-1}\bUtilde^*_{2}$ \Comment{Estimate actuation matrix $\hat{\bB}$.}
		\State $[\bW, \bLambda] \leftarrow \text{EIG}(\bAtilde)$  \Comment{Eigendecomposition of $\bAtilde$.}
		\State
		$\bPhi \leftarrow \bX'\bVtilde\bSigmatilde^{-1}\bUtilde^*_{1}\bUhat\bW $    \Comment{DMD modes of $\bA$.}
		\EndIf
		\EndProcedure
	\end{algorithmic}
Note: If $\blambda_i=0$, then $\bphi_i = \bUtilde_{1}\bUtilde^*_{1}\bUhat\bw_{i} \text{ for step } 13$.
\end{algorithm*}
%% ======================================================================= %%
\subsection{\hspace{-.152in}Dynamic mode decomposition with control}\label{Sec:DMDc}

The dynamic mode decomposition with control (DMDc) method is a critically enabling extension of DMD~\cite{Proctor2016siads}.  
DMDc disambiguates between the underlying dynamics and the effects of actuation, modifying the basic assumption of DMD to include the effect of inputs $\bu_k \in \bR^q$
\begin{align} \label{eqn:4}
\bx_{k+1}=\bA\bx_k+\bB\bu_k
\end{align}
where $\bB \in \bR^{n\times q}$ .  The matrix form of the actuation is
\begin{align} \label{eqn:5}
\bUpsilon=
\begin{bmatrix}
\vline & \vline& &\vline\\
\bu_0&\bu_1 &\hdots&\bu_{m-1}\\
\vline&\vline& &\vline
\end{bmatrix}.
\end{align}
The system can be written in matrix form as:
\begin{align} \label{eqn:6}
\bX'=\bA\bX+\bB\bUpsilon
\end{align}
where each column $\bu_k$ of the input snapshot matrix $\bUpsilon$ is the input at each snapshot in Eq.~\eqref{eqn:5} and the data matrices $\bX,\bX'$ are formulated in the same way as in Eq.~\eqref{Eqn:SnapshotMatrices}.  

A least-squares regression algorithm can once again be used to determine the matrix $\bA$ and its associated DMD modes and eigenvalues.  Two distinguishing cases can be considered, when $\bB$ is known and when $\bB$ is unknown. 
When the input matrix $\bB$ is known, or can be well-estimated, the output is a simple linear combination of states and inputs.  The DMD modes can then be obtained following the procedure of the exact DMD algorithm~\ref{alg DMD} with $\bX'$ replaced with $\bX'-\bB\bUpsilon$.  For an unknown $\bB$ it is possible to compute the DMD modes and an approximation to the matrix $\bB$ via regression~\cite{Proctor2016siads}.   For this case, an augmented matrix containing both the state and input snapshots is constructed
\begin{align} 
\bOmega=  {\begin{bmatrix} \bX \\ \bUpsilon \end{bmatrix}}  \label{Eq:Omega}
\end{align}
along with an augmented matrix containing the two unknown system matrices
\begin{align}
\bG=\begin{bmatrix}\bA & \bB\end{bmatrix}  \, .
\end{align}
The regression problem is then formulated as
\begin{align} \label{eqn:7}
\bX'=\begin{bmatrix} \bA& \bB \end {bmatrix} \begin{bmatrix} \bX\\ \bUpsilon\end{bmatrix}= \bG \bOmega
\end{align}
where $\bOmega \in \bR^{(n+q)\times m}$ is the combination of the state and control snapshots. The DMDc algorithm with an unknown $\bB$ is shown in Algorithm~\ref{alg DMDc}.  Importantly, the SVD now constructs a low-rank representation of the state and input variables, both of which are used to produce approximations of the matrices $\bA$ and $\bB$ through low-rank structures.
Much like the DMD algorithm, DMDc relies on least-squares regression of the data to build a linear model $\bA$ of the state dynamics that is disambiguated from the discovered actuation matrix $\bB$.

\begin{algorithm*}
	\caption{Compressive DMD~\cite{Brunton2015jcd}} \label{alg cDMD}
	\begin{algorithmic}[1]
		\INPUT{Measurements $\bY,\bY'$, measurement matrix $\bC$, sparsifying basis $\bPsi$, and target rank $r$.}
		\OPTIONAL{$\bX,\bX'$.}
		\OUTPUT{cDMD spectrum $\bLambda$ and modes $\hat{\bPhi}$.}
		\Procedure{cDMD}{$\bY,\bY',\bC,r,[\bX,\bX']$}.
		\State $[\bUY, \bSigmaY, \bVY] \leftarrow$ SVD($\bY,r$) \Comment{Truncated $r$-rank SVD of $\bY$.}
		\State $\bAtildeY \leftarrow \bU^*_{\bY}\bY'\bVY\bSigma^{-1}_{\bY}$ \Comment{Low-rank approximation of $\bAY$.}
		\State $[\bWY, \bLambdaY] \leftarrow \text{EIG} (\bAtildeY)$ \Comment{Eigendecomposition of $\bAtildeY$.}
		\If{$\bX$ is known} \Comment{Perform \emph{compressed DMD}.}
		\State $\hat{\bPhi} \leftarrow%\Comment Compute the (truncated $r$-rank) SVD of $\bX$.
			\bX'\bVY\bSigma^{-1}_{\bY}\bWY $ \Comment{Estimate DMD modes of $\bA$.}
		\Else \Comment{Perform \emph{compressed sensing DMD}.}
		\State $\bPhiY \leftarrow \bY'\bVY\bSigma^{-1}_{\bY}\bWY $ \Comment{DMD modes of $\bAY$.}%\Comment Obtain compressed dynamic modes of the operator $\bA$.
		\State
		$\bPhiS \leftarrow$ Compressed Sensing($\bPhiY,\bC,\bPsi$)
		\Comment{Perform $l_1$ minimization on $\bphiY_i$ to solve for $\bphiS_i$.}
		\State
		$\hat{\bPhi} \leftarrow \bPsi\bPhiS$   \Comment{Estimate DMD modes of $\bA$.}
                 \EndIf
		\EndProcedure
	\end{algorithmic}
	Note: If $\blambda_i=0$, then $\bphi_i = \bU\bw_{\bY,i} , \bphi_{\bY,i} = \bUY\bw_{\bY,i}$ for steps $6$ and $8$. 
\end{algorithm*}

%% ======================================================================= %%
\subsection{Compressed sensing and dynamic mode decomposition}\label{Sec:CS}
Another innovation of the DMD algorithm addresses limitations on measurement and acquisition of a dynamical system.  Such limited data acquisition is often imposed by physical constraints, such as data-transfer bandwidth in particle image velocimetry (PIV), or the costs of sensors.  Given the low-rank nature of the 
spatiotemporal structures exhibited in many complex systems, we can utilize ideas from compressed sensing~\cite{Donoho2006ieee,Candes2006ieee,Baraniuk2007ieee} to reconstruct the full high-dimensional state $\bx$ from a small number of measurements.  
Compressive DMD (cDMD) develops a strategy for computing the dynamic mode decomposition from compressed or 
subsampled data~\cite{Brunton2015jcd}.  

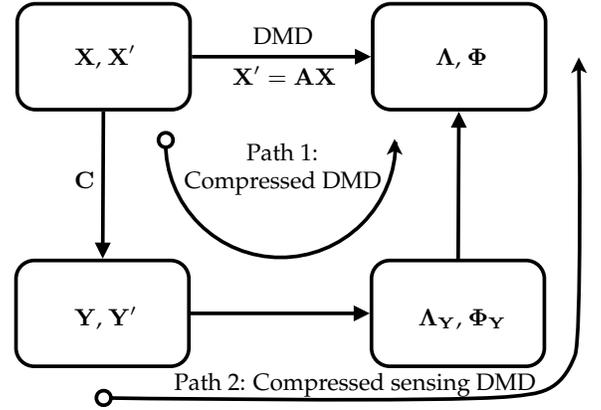
\begin{figure}[t]
\vspace{-.3in}
\begin{center}
\begin{overpic}[width=.5\textwidth]{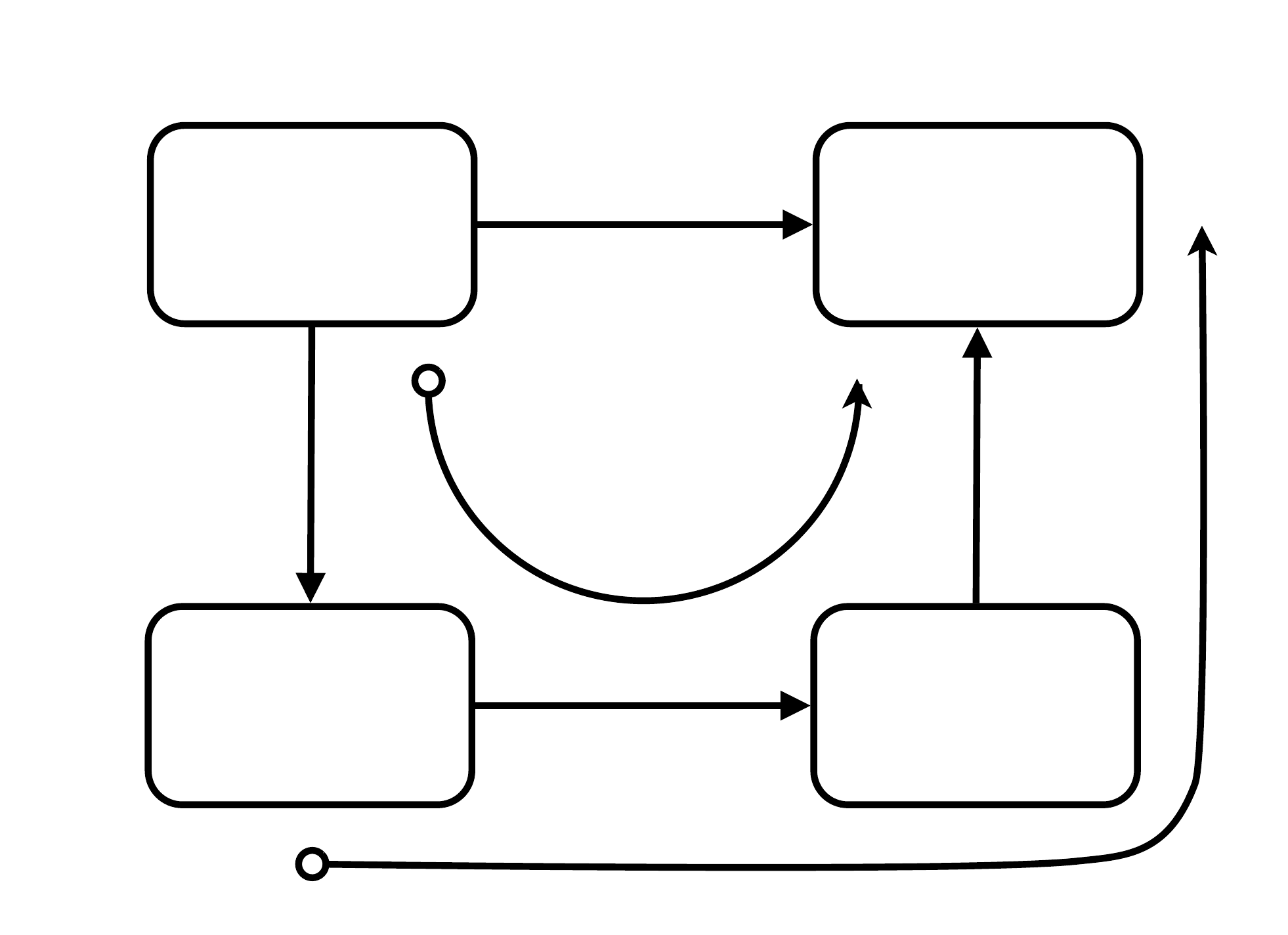}
\small
\put(20,54){$\bX$, $\bX'$}
\put(20,36){$\bC$}
\put(20,16){$\bY$, $\bY'$}
\put(46,57){DMD}
\put(43,51){$\bX'=\bA\bX$}
\put(72.5,54){$\bLambda$, $\bPhi$}
\put(70,16){$\bLambdaY$, $\bPhiY$}
\put(45,40){Path 1:}
\put(36,36){Compressed DMD}
\put(34.5,6.7){Path 2: Compressed sensing DMD}
\end{overpic}
\vspace{-.3in}
\caption{Schematic of DMD and compressive DMD. Path $1$ shows compressed DMD and path $2$ shows compressed sensing DMD (~\cite{Brunton2015jcd}).}\label{fig01}
\vspace{-.3in}
\end{center}
\end{figure}

Consider compressed or subsampled data $\bY$ given by 
\begin{align}
\bY = \bC\bX,
\end{align}
where $\bC$ is a measurement matrix.  
There are two key strategies to cDMD as illustrated in Fig.~\ref{fig01}. 
First, it is possible to reconstruct full-state DMD modes from heavily subsampled or compressed data using compressed sensing. 
This is called compressed sensing DMD, and it is appropriate to use when access to the full state space is not possible due to constraints on physical measurements. 
Second, if full-state snapshots are available, it is possible to first compress the data, perform DMD, and then reconstruct by taking a linear combination of the snapshot data, determined by the DMD on compressed data. 
This is called compressed DMD. 
In this case, it is assumed that one has access to the full high-dimensional state space data.  
The theory for either of these methods relies on relationships between DMD on full-state and compressed data. 
Importantly, when data and modes are sparse in some transform basis, then there is an invariance of DMD to measurement matrices that satisfy the restricted isometry property (RIP) from compressed sensing. 

In addition to the the data matrices $\bX$ and $\bX'$, it is also now required to specify a measurements matrix $\bC$.  These three matrices together are required to execute Algorithm~\ref{alg cDMD}.   In practice, we often consider point measurements so that the rows of $\bC$ are rows of the identity matrix.  For DMD modes that are global in nature, point measurements naturally satisfy the RIP property as they are incoherent with respect to the global modes.  The success of the method, both the compressive-sampling DMD and the cDMD, has been demonstrated by Brunton {\em et al.}~\cite{Brunton2015jcd} to be an effective strategy for subsampling of data and the reconstruction of DMD modes and eigenvalues.

%Compressive sensing references~\cite{Donoho:2006,Candes:2006,Baraniuk:2007}
%
%
%how and why we can expect to do compressed measurements.
%
%cite everybody (CS groups... also cite Bai and Glauser, and Brunton, and Rowley) (especially CS in fluids)

%%%%%%%%%%%%
%%% METHODS
%%%%%%%%%%%%
\section{Compressive system identification}\label{sec:methods}
A major benefit of dynamic mode decomposition is that it provides a physically interpretable and highly extensible linear model framework, which enables the incorporation of actuation inputs and sparse output measurements.  
When combined, these innovations result in the compressive DMD with control (cDMDc) architecture for \emph{compressive system identification}, where low-order models are identified from input--output measurements.  
In contrast to traditional system identification, the reduced states of the cDMDc models may be used to reconstruct the high-dimensional state space via compressed sensing, adding physical interpretability to the models.  
Thus, cDMDc relies on the existence of a few dominant coherent patterns, which in turn facilitates sparse measurements.  

We now consider a general input--output system with high-dimensional state $\bx$:
\begin{subequations} \label{eqn:ssdisc}
\begin{eqnarray}
\bx_{k+1} &= & \bA \bx_k + \bB \bu_k\label{eq:ssdisc:state}\\
\by_k & = & \bC \bx_k\label{eq:ssdisc:output}.
\end{eqnarray}
\end{subequations}
As in DMD, the goal is to obtain the leading eigendecomposition of $\bA$, resulting in a low-order model in terms of a few DMD modes. 
However, now we must account for limited measurements in the output $\by$ and disambiguate internal state dynamics from the effect of actuation $\bu$.  
Writing Eq.~\eqref{eqn:ssdisc} in terms of data matrices yields: %becomes:
\begin{subequations} \label{eqn:8}
\begin{eqnarray}
\bX' &= & \bA \bX + \bB \bUpsilon\\
\bY & = & \bC \bX.
\end{eqnarray}
\end{subequations}

\begin{figure}
\vspace{-.3in}
\begin{center}
\begin{overpic}[width=.5\textwidth]{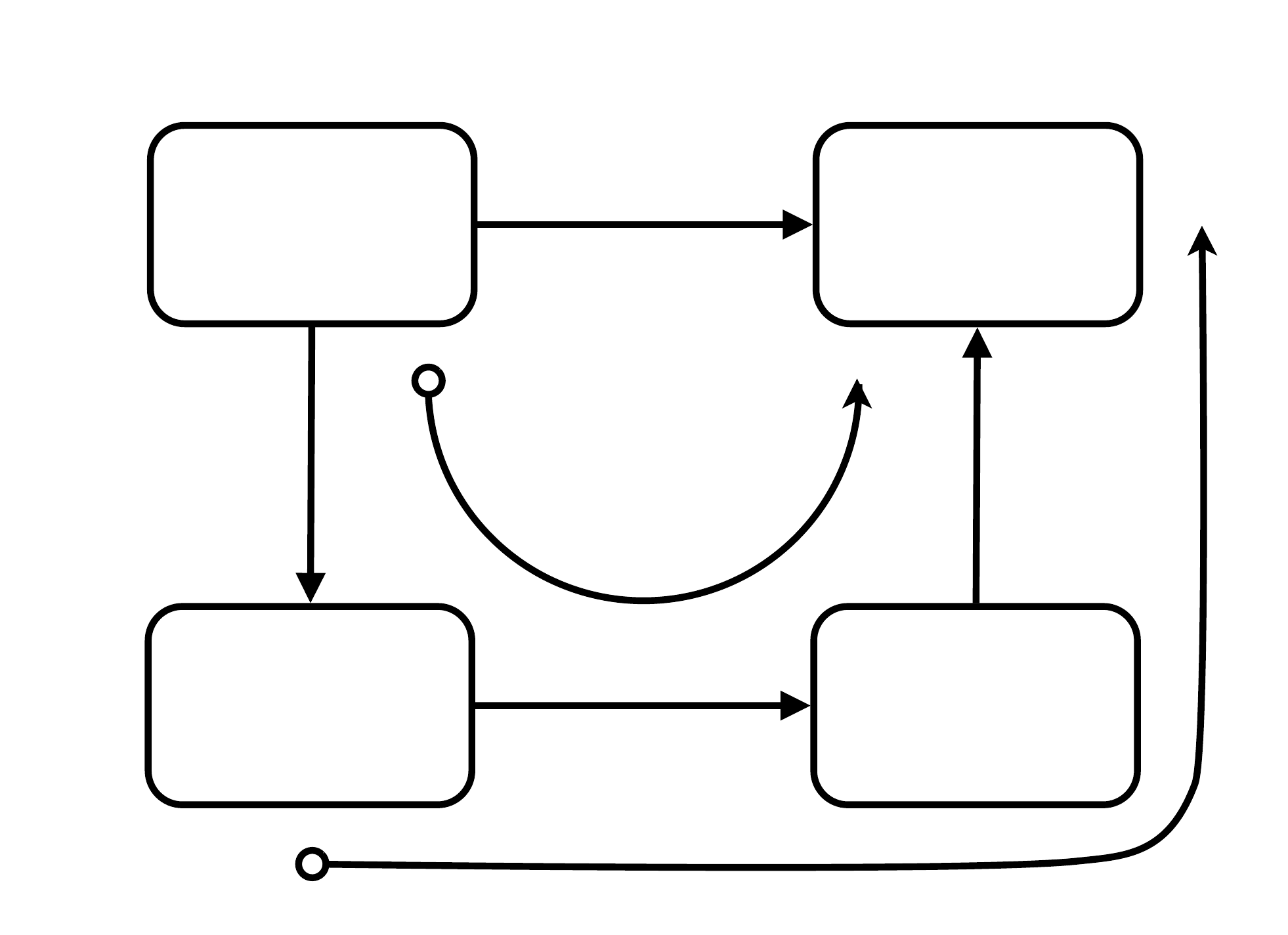} 
\small
\put(14.5,54){$\bX$, $\bX'$, $\bUpsilon$,  $[\bB]$}
\put(20,36){$\bC$}
\put(14.5,16){$\bY$, $\bY'$, $\bUpsilon$, $[\bB]$}
\put(45,57){DMDc}
\put(37.7,51){$\bX'=\bA\bX+\bB\bUpsilon$}
\put(69.5,54){$\bLambda$, $\bPhi, [\bBhat]$}
\put(65,16){$\bLambdaY$, $\bPhiY, [\bBYhat]$}
\put(45,41.5){Path $1$:}
\put(35.,37.5){Compressed DMDc}
\put(33,6.8){Path $2$: Compressed sensing DMDc}
\end{overpic}
\vspace{-.35in}
\caption{Schematic of compressive DMD control. Path $1$ shows compressed DMDc and path $2$ shows compressed sensing DMDc.}\label{sche cDMDc}
\vspace{-.25in}
\end{center}
\end{figure}

Under certain conditions it is possible to apply DMDc to the compressed data $\bY$ and then recover full-state DMD modes via compressed sensing.  
As in the compressive DMD algorithm~\cite{Brunton2015jcd}, if full-state data $\bX$ is available, significant computational savings may be attained by compressing the data and working in the compressed subspace.  
If full-state data is unavailable, it may still be possible to reconstruct full-state modes via convex optimization, exploiting sparsity of the modes in a generic basis, such as Fourier or wavelets.  The cDMDc framework is shown in Fig.~\ref{sche cDMDc}, and is described in algorithm~\ref{alg cDMDc}.  

\begin{algorithm*}
	\caption{Compressive DMD with control} \label{alg cDMDc}
	\begin{algorithmic}[1]
	\INPUT{Measurement snapshot matrices $\bY,\bY'$, sensing matrix $\bC$, input snapshot matrix $\bUpsilon$, and target rank $r,\rtilde$.}
	\OPTIONAL{Full-state data matrices $\bX,\bX'$ and actuation matrix $\bB$.}
	\OUTPUT{cDMDc spectrum $\bLambda$ and modes $\hat{\bPhi}, [\bBhat]$}.  %\Comment{Optional: the estimated $\bB$.}
	\Procedure{cDMDc}{$\bY,\bY',\bC,r,\rtilde,[\bX,\bX',\bB]$}
	\If{$\bX$ is known}
	     \If{$\bB$ is known} 
                 \State $\bLambda,\hat{\bPhi} \leftarrow$ cDMD($\bY,\bY'-\bC\bB\bUpsilon,\bC,\bX,\bX',r$) \Comment{Perform \emph{compressed DMD}.}
	     \Else
		\State $\bLambdaY,\bUhatY,\bUtilde_{\bY,1},\bUtilde_{\bY,2},\bSigmatilde_{\bY},\bVtildeY,\bWY$
		\State \hspace{0.38cm}$\leftarrow$ DMDc($\bY,\bY',\bUpsilon,r,\rtilde, \bX, \bX'-\bB\bUpsilon$)  \Comment{Perform \emph{DMDc}.}
		\State
		$\hat{\bPhi} \leftarrow \bX'\bVtildeY\bSigmatilde^{-1}_{\bY}\bUtilde^*_{\bY,1}\bUhatY\bWY$ \Comment{Estimate DMD modes of $\bA$.}			
		\State
		$\bBhat \leftarrow \bX'\bVtildeY\bSigmatilde^{-1}_{\bY}\bUtilde^*_{\bY,2}$ \Comment{Estimate actuation matrix $\hat{\bB}$.}
	     \EndIf
	\Else
	    \If{$\bB$ is known}
                \State $\bLambda,\hat{\bPhi} \leftarrow$ cDMD($\bY,\bY'-\bC\bB\bUpsilon,\bC,r$) \Comment{Perform \emph{compressed sensing DMD}.}
            \Else
                \State $\bLambdaY,\bPhiY,\bBYhat \leftarrow$ DMDc($\bY,\bY',\bUpsilon,r,\rtilde$) \Comment{Perform \emph{DMDc}.}
                \State $\hat{\bPhi} \leftarrow$ Compressed Sensing($\bPhiY,\bPsi$) \Comment{Estimate DMD modes of $\bA$.}		
	        \State $\bBhat \leftarrow$ Compressed Sensing($\bBYhat,\bPsi$) \Comment{Estimate actuation matrix $\hat{\bB}$.}

	    \EndIf
	\EndIf
	\EndProcedure
	\end{algorithmic}
	Note: If $\lambda_i=0$, then $\bphi_i = \bUtilde\bUtilde^*_{\bY,1}\bUhatY\bw_{\bY,i} \text{ for step } 7.$
\end{algorithm*}

We now prove that when compressed, DMD eigenvectors of the full data $\bX$ become DMD eigenvectors of the compressed data $\bY$.   
This section relies on notation developed above, which is consolidated in the nomenclature.  
Matrices with a subscript $\bY$ are computed on compressed data, and matrices with a tilde are computed from the SVD of the augmented matrix $\bOmega$.  

\begin{assumption} \label{assump1}
The measurement matrix $\bC$ preserves the temporal information in $\bOmega$ so that $\bVtildeY\bVtildeY^*\bVtilde=\bVtilde$. 
This requires the columns of $\bVtilde$ to be in the column space of $\bVtildeY$ and will only be approximately satisfied with measurement noise.
\end{assumption}
\begin{assumption} \label{assump2}
The columns of the full-state output matrix $\bX'$ are in the subspace of the upper left singular vectors $\bUtilde_1$ of the full-state augmented matrix $\bOmega$ so that $\bUtilde_1\bUtilde^*_1\bX' \approx \bX'$. 
\end{assumption}

\begin{lemma} \label{lemma1}
If Assumption~\ref{assump1} holds, we have an identity $\bVtilde\bSigmatilde^{-1} = \bVtildeY\bSigmatilde^{-1}_{\bY}\bUtilde_{\bY,1}^*\bC\bUtilde_1$, similar to that derived in ~\cite{Brunton2015jcd}:
\begin{proof}
~\\
\vspace{-.45in}
\begin{align*}
\bY&=\bC\bX \nonumber \\
\bUtilde_{\bY,1}\bSigmatildeY\bVtildeY^* &= \bC\bUtilde_1\bSigmatilde\bVtilde^* \nonumber \\
%\bC\bUtilde_1\bSigmatilde\bVtilde^* &\approx \bUtilde_{\bY,1}\bSigmatildeY\bVtildeY^* \nonumber \\
\bVtilde^*_{\bY}\bVtilde\bSigmatilde^{-1} &= \bSigmatilde^{-1}_{\bY}\bUtilde_{\bY,1}^*\bC\bUtilde_1  \nonumber \\
\bVtildeY\bVtildeY^*\bVtilde\bSigmatilde^{-1} &= \bVtildeY\bSigmatilde^{-1}_{\bY}\bUtilde_{\bY,1}^*\bC\bUtilde _1 \nonumber \\
\bVtilde\bSigmatilde^{-1} &= \bVtildeY\bSigmatilde^{-1}_{\bY}\bUtilde_{\bY,1}^*\bC\bUtilde_1.
\end{align*}
~\vspace{-.475in}\\
\vspace{-.1in}
\end{proof}
\end{lemma}

The next theorem uses the following definitions of $\bA$ and $\bAY$ from the DMDc and cDMDc algorithms:
\begin{align*}
\bA& = \bX' \bVtilde\bSigmatilde^{-1}\bUtilde^*_1\\
\bAY& = \bY'\bVtildeY\bSigmatilde^{-1}_{\bY}\bUtilde_{\bY,1}^*.
\end{align*}

\begin{theorem} \label{theorem1}
If Assumption~\ref{assump1} and~\ref{assump2} hold, then
\begin{align}
\bC\bA\bX'=\bAY\bC\bX'.
\end{align}
\begin{proof}
Using Lemma~\ref{lemma1}, we have
\begin{align*}
\bC\bA\bX'&=\bC(\bX'\bVtilde\bSigmatilde^{-1}\bUtilde^*_1)\bX' \nonumber \\
&=\bY'(\bVtildeY\bSigmatilde^{-1}_{\bY}\bUtilde_{\bY,1}^*\bC\bUtilde_1)\bUtilde^*_1\bX' \nonumber \\
&=\bAY\bC\bUtilde_1\bUtilde^*_1\bX'\nonumber\\
&=\bAY\bC\bX'.
\end{align*}
~\vspace{-.475in}\\
\vspace{-.1in}
\end{proof}
\end{theorem}

Thus, the compression matrix $\bC$ commutes with the dynamics in $\bA$, when applied to data $\bX'$.  
We use this to obtain the central result: compressed dynamic modes of the full data are dynamic modes of the compressed data.
\begin{theorem} \label{theorem3}
If Assumptions~\ref{assump1} and~\ref{assump2} hold, then compressing a full-state DMDc eigenvector $\bphi$ yields a DMDc eigenvector of the compressed data with the same eigenvalue:
\begin{align}
\bAY\bC\bphi = \lambda \bC\bphi.
\end{align}
\begin{proof}
~\\
\vspace{-.5in}
\begin{align*}
\bAY\bC\bphi &= \bAY\bC\bX'\bVtilde\bSigmatilde^{-1}\bUtilde_1^*\bUhat\bw\nonumber\\
&=\bC\bA\bX'\bVtilde\bSigmatilde^{-1}\bUtilde_1^*\bUhat\bw \nonumber\\
&=\bC\bA\bphi \nonumber\\
&=\lambda\bC\bphi.  
\end{align*}
~\vspace{-.475in}\\
\vspace{-.1in}
\end{proof}
\end{theorem}

If $\bC$ is chosen poorly so that $\bphi$ is in its nullspace, then Theorem~\ref{theorem3} applies trivially.  This theorem does not guarantee that every eigenvector of $\bAY$ is a compressed eigenvector of $\bA$, although under reasonable assumptions the dominant eigenvalues of $\bAY$ will approximate those of $\bA$, as shown in~\cite{Brunton2015jcd}.  
We then have
\begin{align}
\bphi_{\bY} = \bC\bphi_{\bX} = \bC\bPsi\bphi_{\bS},
\end{align}
where $\bphi_{\bS}$ is the sparse representation of $\bphi_{\bX}$ in a universal basis $\bPsi$, such as a Fourier or wavelet basis.  
Thus, it is possible to recover $\bphi_{\bS}$, and hence $\bphi_{\bX}$, from compressed DMD modes $\bphi_{\bY}$, given enough incoherent measurements~\cite{Brunton2015jcd}.  

\vspace{-.1in}
\paragraph{Compressed recovery of the actuation matrix.}
We now establish a similar relationship between the full-state actuation matrix $\bB$ and the compressed matrix $\bB_\bY$.  

\begin{assumption} \label{assump3}
The columns of $\bUtilde_{\bY,2}$ are spanned by those of $\bUtilde_2$, so $\bUtilde_2\bUtilde_2^*\bUtilde_{\bY,2}=\bUtilde_{\bY,2}$ and $\bUtilde_{\bY,2}^*\bUtilde_2\bUtilde_2^*=\bUtilde_{\bY,2}^*$. 
\end{assumption}

%\begin{assumption} \label{assump3}
%The columns of the control input matrix $\bUpsilon$ are in the subspace of the lower left singular vectors $\bUtilde_2$ of the full-state augmented matrix $\bOmega$ so that $\bUtilde_2\bUtilde^*_2\bUpsilon \approx \bUpsilon$. 
%\end{assumption}

\begin{lemma} \label{lemma2}
We have an identity $\bVtilde\bSigmatilde^{-1} = \bVtildeY\bSigmatilde^{-1}_{\bY}\bUtilde_{\bY,2}^*\bUtilde_2$.
\begin{proof}
Expanding $\bUpsilon$ in the SVD bases of $\bOmega_\bY$ and $\bOmega$, we find:
%From DMDc [Alg.~\ref{alg DMDc}] on $\bX$ and cDMDc [Alg.~\ref{alg cDMDc}] on $\bY$,  we have $\bUpsilon$ reconstructed as
\begin{align*}
\bUtilde_{\bY,2}\bSigmatildeY\bVtildeY^* &= \bUtilde_2\bSigmatilde\bVtilde^* \\
\bVtilde^*_{\bY}\bVtilde\bSigmatilde^{-1}&=\bSigmatilde^{-1}_{\bY}\bUtilde_{\bY,2}^*\bUtilde_2 \\  
\bVtildeY\bVtildeY^*\bVtilde\bSigmatilde^{-1}&=\bVtildeY\bSigmatilde^{-1}_{\bY}\bUtilde_{\bY,2}^*\bUtilde _2 \\
\bVtilde\bSigmatilde^{-1} &= \bVtildeY\bSigmatilde^{-1}_{\bY}\bUtilde_{\bY,2}^*\bUtilde_2.
\end{align*}
~\vspace{-.475in}\\
\vspace{-.1in}
\end{proof}
\end{lemma}

The next theorem uses the following definitions of $\bB$ and $\bBY$ from the DMDc and cDMDc algorithms:
\begin{align*}
\bB& = \bX' \bVtilde\bSigmatilde^{-1}\bUtilde^*_2\\
\bBY& = \bY'\bVtildeY\bSigmatilde^{-1}_{\bY}\bUtilde_{\bY,2}^*.
\end{align*}
%

%\begin{theorem} \label{theorem2}
%If Assumption~\ref{assump1} and~\ref{assump3} hold, then
%\begin{align}
%\bC\bB\bUpsilon=\bBY\bUpsilon.
%\end{align}
%\begin{proof}
%Using Lemma~\ref{lemma2}, we have
%\begin{align}
%\bC\bB\bUpsilon &=\bC(\bX'\bVtilde\bSigmatilde^{-1}\bUtilde^*_2)\bUpsilon \nonumber \\
%&=\bY'(\bVtildeY\bSigmatilde^{-1}_{\bY}\bUtilde_{\bY,2}^*\bUtilde_2)\bUtilde^*_2\bUpsilon \nonumber \\
%&=\bBY\bUtilde_2\bUtilde^*_2\bUpsilon\nonumber\\
%&=\bBY\bUpsilon.
%\end{align}
%\end{proof}
%\end{theorem}
\begin{theorem} \label{theorem2}
If Assumptions~\ref{assump1} and~\ref{assump3} hold, then
\begin{align}
\bC\bB=\bBY.
\end{align}
\begin{proof}
Using Lemma~\ref{lemma2}, we have
\begin{align*}
\bC\bB &=\bC\bX'\bVtilde\bSigmatilde^{-1}\bUtilde^*_2 \nonumber \\
&=\bY'\bVtildeY\bSigmatilde^{-1}_{\bY}\bUtilde_{\bY,2}^*\bUtilde_2\bUtilde^*_2 \nonumber \\
&=\bY'\bVtildeY\bSigmatilde^{-1}_{\bY}\bUtilde_{\bY,2}^*\nonumber\\
&=\bBY.
\end{align*}
~\vspace{-.475in}\\
\vspace{-.1in}
\end{proof}
\end{theorem}

Therefore, we can first compute the DMD modes $\bPhiY$ and the compressed actuation matrix $\bBY$ and then reconstruct the full-state $\bPhi$ and $\bB$ through compressed sensing.

\subsection{Relationship to system identification}

The result from Theorem~\ref{theorem1} above also carries over to general impulse response parameters in the following theorem. 
These impulse response parameters, also known as \emph{Markov parameters}, are used extensively in system identification, for example to construct Hankel matrices in the eigensystem realization algorithm (ERA)~\cite{Juang1985jgcd}.  

\begin{theorem} \label{theorem4}
If Assumptions~\ref{assump1} and~\ref{assump2} hold, then
\begin{align}
\bC\bA\bB=\bAY\bC\bB.
\end{align}
\begin{proof}
Using Lemma~\ref{lemma1}, we have
\begin{align*}
\bC\bA\bB&=\bC(\bX'\bVtilde\bSigmatilde^{-1}\bUtilde^*_1)\bB \nonumber \\
&=\bY'(\bVtildeY\bSigmatilde^{-1}_{\bY}\bUtilde_{\bY,1}^*\bC\bUtilde_1)\bUtilde^*_1\bB \nonumber \\
&=\bAY\bC{\bUtilde_1\bUtilde^*_1\bX'}\bVtilde\bSigmatilde^{-1}\bUtilde^*_2\nonumber\\
&=\bAY\bC\bX'\bVtilde\bSigmatilde^{-1}\bUtilde^*_2 \nonumber \\
&=\bAY\bC\bB.
\end{align*}
~\vspace{-.45in}\\
%\vspace{-.1in}
\end{proof}
\end{theorem}

\begin{corollary}\label{Corr1}
Theorem~\ref{theorem4} can be expanded to further steps in dynamics, for $k\in\mathbb{Z}^+$, such as:
\begin{align}
\bC\bA^k\bB=\bAY^k\bC\bB = \bAY^k\bBY.
\end{align}
\end{corollary}

This theorem and corollary establish a surprising result: under certain conditions the iterative dynamics in the impulse response parameters commute with the measurement matrix.  
Impulse response parameters are a cornerstone of subspace system identification methods, such as ERA, and the above results simplify the dynamics.  

Corollary~\ref{Corr1} also has implications for the controllability of the projected and full-state systems.  Given the controllability matrix 
\begin{align} \label{eqn:ctrb}
{\mathcal {C}} = \begin{bmatrix}\bB &\bA\bB &\cdots   \bA^{n-1}\bB  \end{bmatrix}
\end{align}
we have the following relationship:
\begin{subequations}\label{eqn:ctrb2}
\begin{align} 
\bC\,{\mathcal {C}} &= \bC\begin{bmatrix}\bB &\bA\bB &\cdots   \bA^{n-1}\bB  \end{bmatrix}\\
&=  \begin{bmatrix} \bBY& \bA_{\bY}\bBY & \cdots  &\bA^{n-1}_{\bY}\bBY \end{bmatrix} = \mathcal{C}_{\bY}.
\end{align}
\end{subequations}

%%The full-state and compressed measurements are further related by the following
%%\begin{align} \label{eqn:framework}
%%\bY' &= \bC\bX' \nonumber \\ 
%%	&= \bC\bA\bX + \bC\bB\bUpsilon \nonumber \\ 
%%	 &\approx \bAY \bY + \bBY \bUpsilon.
%%\end{align}
%%The dynamics of the compressed system has been discovered above. 
%For the $k$-th snapshot of the discrete-time dynamical system as in Eq.~\eqref{eqn:framework}, we have
%\begin{align*}
%%\by_k &= \bC\bA^{k}\bx_0 + \bC(\bB\bu_{k-1}+\bA\bB\bu_{k-2}+\cdots+\bA^{k-1}\bB\bu_0) \nonumber \\ 
%	  &= \bC\bA^{k}\bx_0 + \bC\begin{bmatrix}\bB &\bA\bB &\cdots   \bA^{k-1}\bB  \end{bmatrix} \begin{bmatrix}\bu_{k-1} \\ \bu_{k-2} \\ \vdots  \\ \bu_0 \end{bmatrix} \\ 
%%	  &= \underbrace{\bC\bPhi}_{\bPhiY}\bLambda^{k}\bb + \begin{bmatrix}\bA^{k-1}_{\bY}\underbrace{\bC\bB}_{\bBY} &\bA^{k-2}_{\bY}\underbrace{\bC\bB}_{\bBY}  &\cdots  &\underbrace{\bC\bB}_{\bBY}\end{bmatrix} \begin{bmatrix}\bu_0 \\ \bu_1  \\ \vdots  \\ \bu_{k-1} \end{bmatrix} \nonumber \\ 
%	  &= \bAY^{k}\by_0 + \begin{bmatrix} \bBY& \bA_{\bY}\bBY & \cdots  &\bA^{k-1}_{\bY}\bBY \end{bmatrix} \begin{bmatrix}\bu_{k-1} \\ \bu_{k-2} \\ \vdots  \\ \bu_0 \end{bmatrix}. \\
%\end{align*}
%which gives $\bC{\mathcal {C}} = {\mathcal {C}}_{\bY}$.

Therefore, if the full-state systems is controllable, i.e. the controllability matrix ${\mathcal {C}}$ is full rank,
and the compressed measurement matrix $\bC$ is not in the null space of Eq.~\eqref{eqn:ctrb},
%\begin{align} \label{eqn:ctrb}
%{\mathcal {C}} = \begin{bmatrix}\bB &\bA\bB &\cdots   \bA^{k-1}\bB  \end{bmatrix}
%\end{align}
then the compressed system is also controllable. 
The controllability is preserved after compression under certain conditions. 
The observability property regarding the compressed sensing framework was discussed in~\cite{Wakin2010IEEE}.

\subsection{Computational considerations}

Similar to cDMD, there are two main paths presented in compressive DMD with control, depending on the availability of the full-state data matrix $\bX$. 
Specifically, we refer to the first path as compressed DMDc (Path $1$ in Figure \ref{sche cDMDc}) if $\bX$ is known and the second path as compressed sensing DMDc (Path $2$ in Figure \ref{sche cDMDc}) if $\bX$ is only partially known due to the lack of full state measurements:
\begin{enumerate}
\item If the full-state measurements $\bX$ are available, compressed DMDc is advantageous as the expensive calculations are performed on the compressed data $\bY$ and the full-state modes are obtained by linearly combining the snapshots of $\bX$. 
\item Without access to full-state measurements $\bX$, compressed sensing DMDc extracts the inherent dynamics from sub-sampled/compressed data in the matrix $\bY$, disambiguating the effect of actuation.  
Full-state modes may then be recovered, under the standard conditions of compressed sensing. 
\end{enumerate}

In addition, two approaches are considered in each path considering the prior knowledge of $\bB$. Generally, the cDMDc algorithm can be simplified as a corrected cDMD algorithm if $\bB$ is known. Otherwise, DMDc is utilized to extract the underlying dynamics from the compressed states $\bY, \bY'$ and then  the modes $\bPhi$ and actuation matrix $\bB$ are reconstructed by either projecting onto the full states $\bX'$ or through compressed sensing. 

%\algblock{If}{EndIf}
%\algcblock[Else]{If}{Else}{EndIf}
%\algcblock{If}{Else}{EndIf}
%\algblock{Begin}{End}

In Algorithm \ref{alg cDMDc}, these four scenarios are discussed.
When both $\bX$ and $\bB$ are known, the spectrum $\bLambdaY$ and modes $\bPhi$ are obtained by performing compressed DMD on the pre-compressed data $\bY$ and the shifted matrix correcting for the effect of control $\bY'-\bC\bB\bUpsilon$.
When $\bX$ is known and $\bB$ is unknown, DMDc is computed on $\bY, \bY'$ and the dynamic modes $\bPhi$ and actuation matrix $\bB$ are reconstructed as a linear combination of the full-state data $\bX'$.  
When $\bX$ is only partially known and $\bB$ is known, compressed sensing DMD is performed on $\bY$ and $\bY'-\bC\bB\bUpsilon$ and the full-state modes $\bPhi$ are reconstructed from the compressed $\bPhiY$.
When both $\bX$ and $\bB$ are unavailable, DMDc is computed on $\bY, \bY'$ and the dynamic modes $\bPhi$ and actuation matrix $\bB$ are reconstructed using compressed sensing. 

%%%%%%%%%%%%
%%% RESULTS
%%%%%%%%%%%%

\begin{figure*}
\begin{center}
\includegraphics[width=\textwidth]{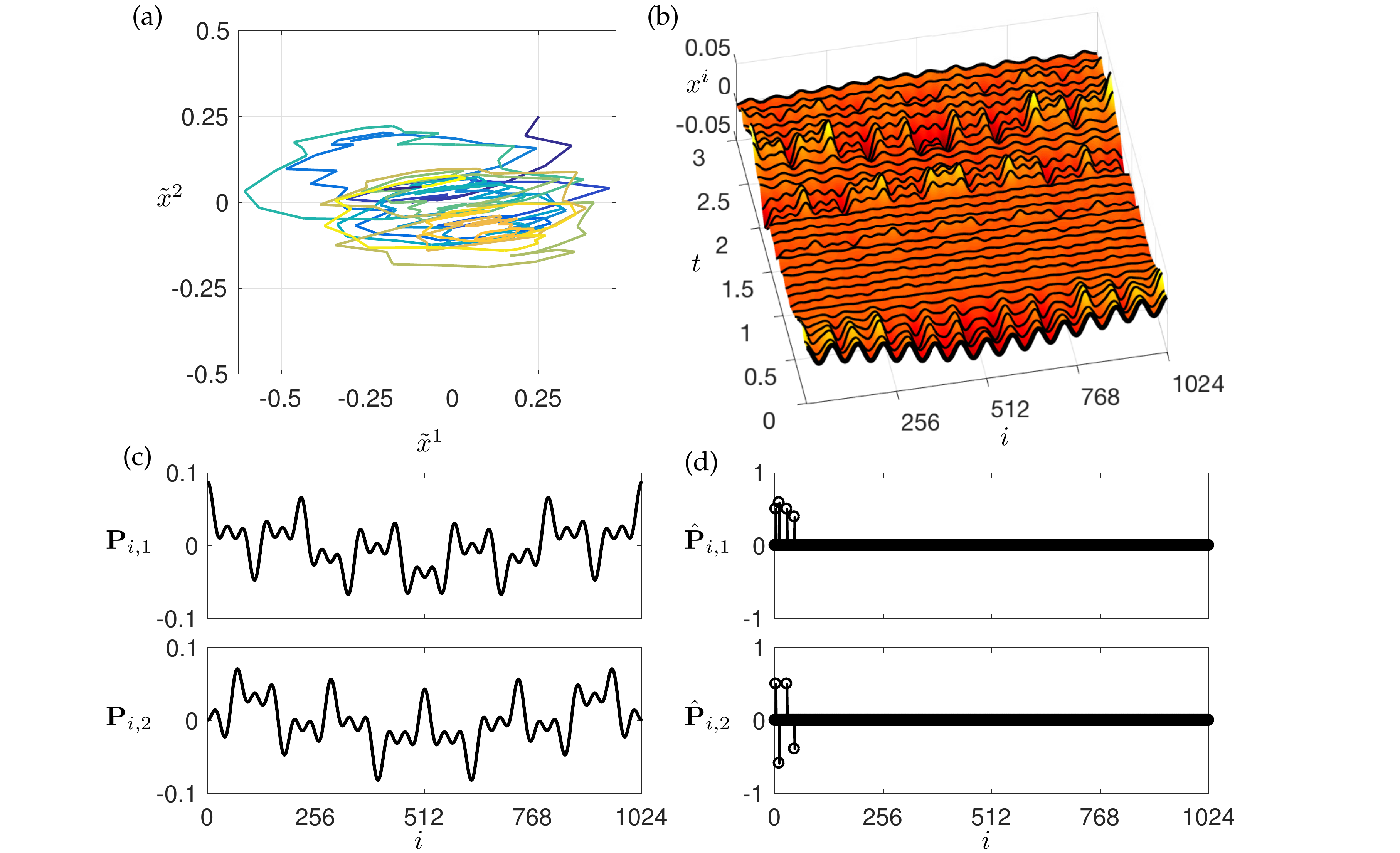}
\caption{Two-dimensional system with known dynamics: (a) phase portrait (color denotes the progression of time), (b) x-t diagram of inflated high-dimensional system, $\bX \in \bR^{1024\times301}$ (first 31 snapshots shown), (c) two orthogonal modes of $\bP$ and (d) DCT coefficients of the two modes of $\bP$.} \label{fig03}
\end{center}
\vspace{-.1in}
\end{figure*}

\section{Results}\label{sec:results}

In this section, we present two numerical experiments that illustrate compressive DMD with control. 
In the first example, we investigate a spatially high-dimensional system that is lifted from a two-dimensional system with known dynamics and random control input. 
In the second example, we model the vorticity field of a fluid flow downstream of a pitching plate at low Reynolds number.  
In both examples, we investigate the effectiveness of different random measurement matrices, compression ratios, and actuation input vectors.  

%In the compressed sensing DMDc approach, we use discrete cosine transform (DCT) matrix as the sparsifying basis $\bPsi$. Specifically, $\bPsi$ is one-dimensional DCT matrix in the first example and two-dimensional DCT2 matrix in the second example.

\subsection{Stochastically forced linear system}\label{sec:ex1}
This experiment is designed to test the compressive DMD with control framework on an example where the low-rank dynamics are known.  
Thus, it is possible to directly compare the true eigenvalue spectrum and spatiotemporal modes with those obtained via compressed DMDc and compressed sensing DMDc. 

The state matrix $\bAtilde$ and input matrix $\bBtilde$ are designed to yield a  stable, controllable system:  
%\begin{subequations}\label{Eq:ToySS}
\begin{align}\label{Eq:ToySS1}
\hspace{-.1in}\begin{bmatrix}\tilde{x}^1\\ \tilde{x}^2\end{bmatrix}_{k+1} &= 
\underbrace{ \begin{bmatrix} 0.9 & 0.2 \\ -0.1 & 0.9 \end{bmatrix}}_{\bAtilde}
\begin{bmatrix}\tilde{x}^1\\ \tilde{x}^2\end{bmatrix}_{k} + 
\underbrace{ \begin{bmatrix} 0.1 \\ 0.01 \end{bmatrix}}_{\bBtilde}u_k.
\end{align}
%\end{subequations}
The dynamics are excited via Gaussian random input excitation in $u_k$, starting from an initial condition $\bx_0 = \begin{bmatrix} 0.25 & 0.25 \end{bmatrix}^T$. 
The system is integrated from $t=0$ to $t=30$ with a time-step of $\Delta t = 0.1$, resulting in $301$ snapshots. 
A sample trajectory of the low-dimensional system is shown in Fig.~\ref{fig03}~(a).   

\begin{table*}
\begin{center}
\setlength{\extrarowheight}{0.2cm}
\begin{tabular}{|c|c|c|c|c|}%{lx{2cm} |x{2cm} |x{2cm} |x{2cm} |x{2cm}|}%
\hline\toprule
\multicolumn{2}{|c|}{} & DMDc & c-DMDc & cs-DMDc \\ \hline
\midrule
\multirow{3}{*}{$\bB$ known} & $\bC$-type $1$ & 3.631e-13 & 3.627e-13 & 3.627e-13\\ \cline{2-5}
	& $\bC$-type $2$ & 3.631 e-13 & 3.443 e-13 &	3.443 e-13 \\ \cline{2-5}
	& $\bC$-type $3$ & 3.631e-13	&  4.210e-13 & 4.210e-13\\ \hline
\midrule
\multirow{3}{*}{$\bB$ unknown} & $\bC$-type $1$ & \diag{.1em}{2.2cm}{4.481e-13}{1.758e-16}& \diag{.1em}{2.2cm}{5.396e-13}{4.840e-17}& \diag{.1em}{2.2cm}{5.926e-13}{9.558e-17}\\[5pt] \cline{2-5}
	& $\bC$-type $2$ & \diag{.1em}{2.2cm}{4.481e-13}{1.758e-16}	& \diag{.1em}{2.2cm}{4.159e-13}{4.731e-17} &	\diag{.1em}{2.2cm}{3.872e-13}{2.845e-16} \\[5pt]  \cline{2-5}
	& $\bC$-type $3$ & \diag{.1em}{2.2cm}{4.481e-13}{1.758e-16}	& \diag{.1em}{2.2cm}{4.322e-13}{3.022e-16} & \diag{.1em}{2.2cm}{5.691e-13}{2.785e-16}	\\[5pt]  \hline
\bottomrule
\end{tabular} 
\caption{Normalized error of $\|\bPhi-\hat{\bPhi}\|_F$ and $\|\bB - \hat{\bB}\|_2$ using DMDc, compressed DMDc, and compressed sensing DMDc. Three types of compression are shown: uniform random projections ($\bC$-type $1$), Gaussian random projections ($\bC$-type $2$) and single pixel measurements ($\bC$-type $3$). When $\bB$ is unknown, the error of the estimated $\hat{\bB}$ is given in the upper triangle, and the error of $\hat{\bPhi}$ is given in the lower triangle.  In all cases, $\bB$ is a linear combination of columns of $\bP$.  } \label{table:error}
\end{center} 
%\vspace{-0.2in}
\end{table*}

To inflate the state dimension, we associate each of the two states $\tilde{x}^1$ and $\tilde{x}^2$ with a high-dimensional mode that is sparse in the spatial wavenumber domain.  
These modes, shown in Fig.~\ref{fig03}~(c), are given by the columns of $\bP\in\mathbb{R}^{1024\times 2}$, where each column is constructed to have $K=4$ non-zero elements in the discrete cosine transform (DCT) basis.  
The nonzero DCT coefficients of each mode have the same wave number with different magnitudes, as shown in Fig.~\ref{fig03}~(d). 
Thus, it is possible to use the DCT matrix as the sparsifying basis for compressed sensing.  
With the spatial modes in $\bP$, it is possible to lift the low-dimensional state $\bxtilde_k$ to a high-dimensional state $\bx_k = \bP\bxtilde_k$. 
The lifted state is shown in Fig.~\ref{fig03}~(b) for the first $31$ snapshots, from $t=0$ to $t=3$. 
Note that the high-dimensional actuation vector $\bB$ is chosen to be in the span of the columns of $\bP$, i.e. $\bB = \bP \tilde{\bB}$, so that the actuation only excites low-dimensional dynamics; other types of actuation will be examined further in Fig.~\ref{FIG:ACTUATION}.

\begin{figure}
\begin{center}
\includegraphics[width=.5\textwidth]{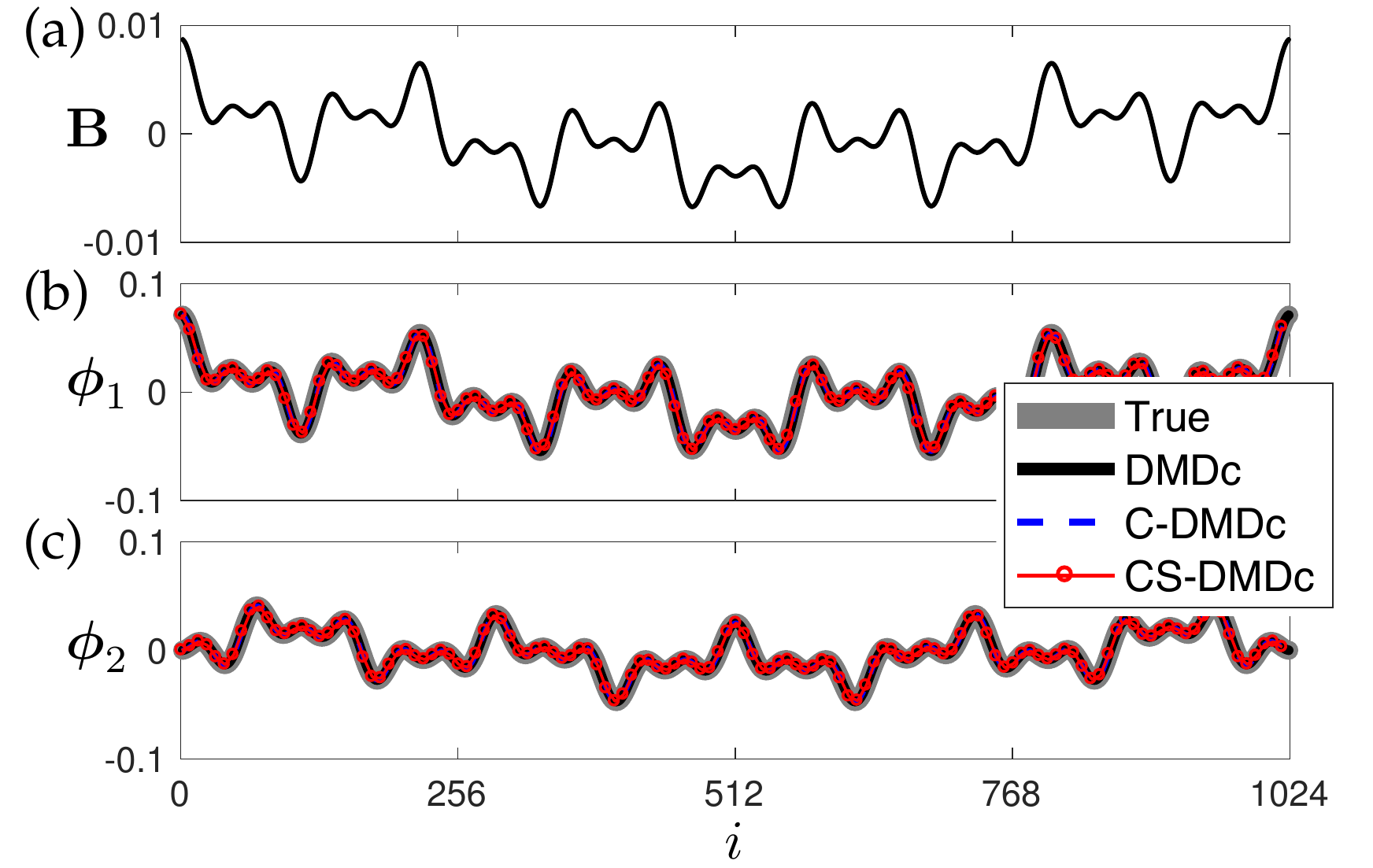}
\vspace{-.2in}
\caption{Reconstruction of (a) actuation vector $\bB$, (b)-(c) modes from DMDc, compressed DMDc and compressed sensing DMDc. For compressive DMD, $128$ Gaussian random measurements are collected from $1024$ state dimensions.}\label{FIG:Ex1:Modes}
\vspace{-.1in}
\end{center}
\end{figure}

To investigate the proposed cDMDc algorithm, we now consider compressed measurements $\by$ given by Eq.~\eqref{eq:ssdisc:output}.  
Specifically, the compression matrix $\bC$ can be built with entries drawn from uniform ($\bC$-type 1), Gaussian ($\bC$-type 2) or Bernoulli ($\bC$-type 3) distributions. 
Fig.~\ref{FIG:Ex1:Modes} shows the DMDc mode reconstruction using compressed DMDc (Path 1 in Fig.~\ref{sche cDMDc}) and compressed sensing DMDc (Path 2 in Fig.~\ref{sche cDMDc}) for $p=128$ Gaussian measurements.  
In both cases, it is assumed that the high-dimensional actuation input vector ${\bB}$ is known, and the reconstructed modes faithfully reproduce the true coherent structures of the underlying system (i.e., the DMD modes are a linear combination of the columns of $\bP$).   
The results remain unchanged when $\bB$ is unknown and must also be reconstructed. 

\begin{figure*}
\vspace{-.15in}
\begin{center}
\includegraphics[width=.975\textwidth]{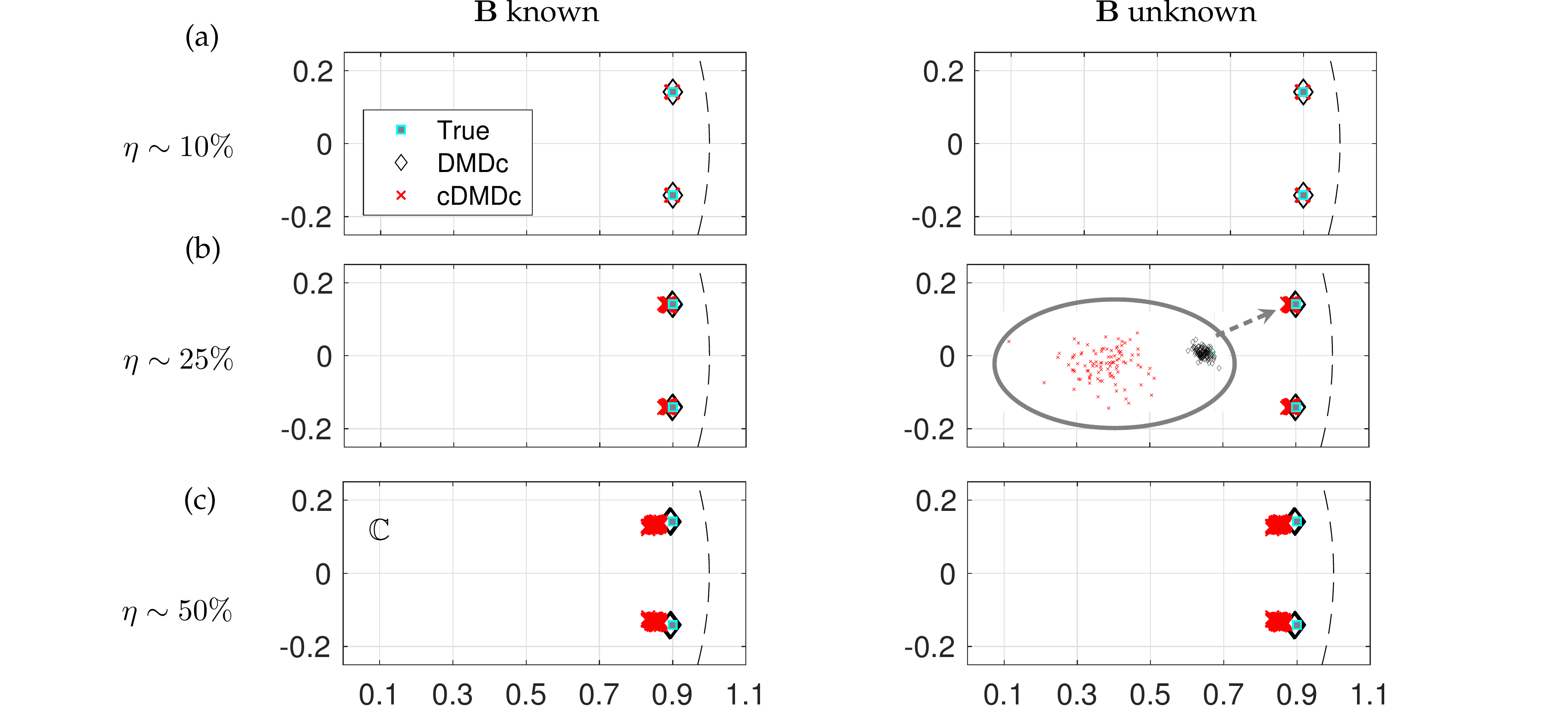}
\vspace{-.1in}
\caption{
	Noise dependency of DMD spectrum for the true system, DMDc and cDMDc based on $100$ different noise realizations.
	Rows correspond to different noise levels $\eta \in\{ 0.1,0.25,0.5 \}$ with $\sigma_{noise} = \eta\, \mathrm{max}(\sigma_i)$, where $\sigma_i$ denotes the standard deviation of each spatial measurement.}
\label{FIG:NOISE}
\end{center}
\end{figure*}

\begin{figure*}
	\centering
	\includegraphics[width=.975\textwidth]{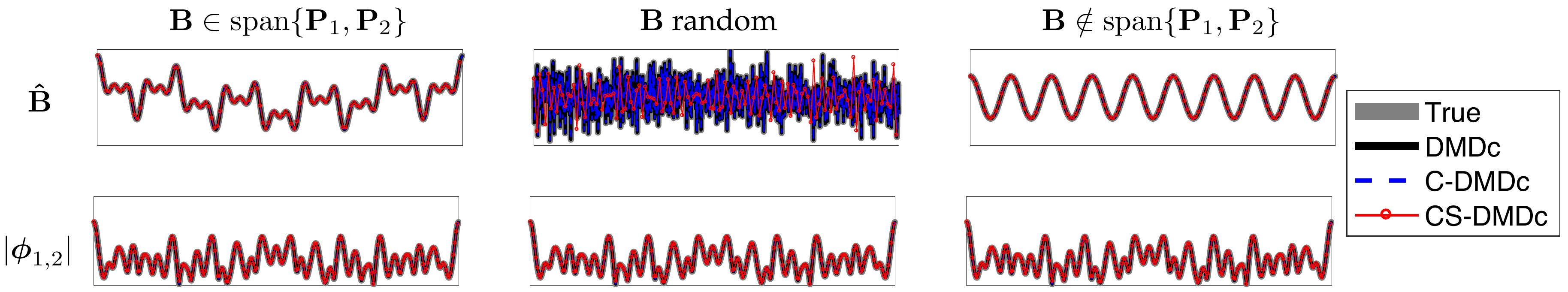}
%\vspace{-.1in}
	\caption{Estimated $\hat\bB$ and mode magnitudes for different choices of $\bB$ for DMDc, compressed DMDc and compressed sensing DMDc. For compressive DMDc, the same number of measurements is employed as for Fig.~\ref{FIG:Ex1:Modes}.}\label{FIG:ACTUATION}
\end{figure*}

Table~\ref{table:error} shows the error in the reconstructed eigenfunctions $\hat\bPhi$ and estimated actuation vector $\hat{\bB}$ (when unknown) for DMDc, compressed DMDc, and compressed sensing DMDc, compared against the true values.  
In addition, we investigate the effect of different sensing strategies discussed above.  %, where $\bC$ is given by either uniform random projections, Gaussian random projections, or by single pixel measurements.  
In all cases, the reconstruction error is small, as no noise is added to the simulated data.  

Compressed sensing DMDc is able to uncover the underlying dynamics and spatio-temporal modes from noiseless subsampled data, relaxing the requirement of high-dimensional measurements.  
However, in realistic experimental conditions, measurement noise will always be present and is known to effect DMD computations.  
Figure~\ref{FIG:NOISE} shows the performance of DMDc and compressed DMDc for varying levels of measurement noise, averaged over 100 different noise realizations in each case.  
Note that compressed sensing DMDc and compressed DMDc have identical eigenvalues, as both methods compute the DMD spectrum from the same compressed data.  
For moderate noise levels, the compressed DMDc algorithm provides reasonably accurate eigenvalues, although they are less accurate than those predicted by DMDc.  
As the noise intensity is increased to greater than 50\%, both the DMDc and cDMDc eigenvalues begin to deviate, and in some cases, complex conjugate eigenvalues are miscomputed as purely real eigenvalues.  
Although the effect of noise is exacerbated by compression, sensitivity of DMD eigenvalues with measurement noise is a known issue, and has been extensively studied and characterized~\cite{Bagheri2014pof,Hemati2015biasing,Dawson2016ef}.  
There are a number of algorithmic extensions that improve the eigenvalue prediction with noise, including using the total least squares~\cite{Hemati2015biasing}, a forward-backward symmetrizing algorithm~\cite{Dawson2016ef}, subspace DMD~\cite{Takeishi2017subspace}, Bayesian DMD~\cite{Takeishi2017JCAI}, or an optimal DMD based on variable projection methods~\cite{Askham2017arxiv}.  
Each of these methods may be effectively combined with the proposed cDMDc architecture to yield more accurate eigenvalues.  

\subsubsection{Effect of Actuation}

Finally, we investigate the performance of cDMDc for different choices of the actuation vector $\bB$ in Fig.~\ref{FIG:ACTUATION}.  
In addition to the cases presented so far, where $\bB$ is in the subspace of $\bP$, we now explore  scenarios when $\bB$ is randomly generated or in the complementary subspace of $\bP$. 
%For all types of input $\bB$, and regardless of whether or not $\bB$ is known, compressed DMDc and compressed sensing DMDc both accurately identify the correct DMDc modes $\bphi_1$ and $\bphi_2$.  
For all types of actuation $\bB$, and regardless of whether or not $\bB$ is known, compressed DMDc and compressed sensing DMDc both accurately identify the true modes $\bphi_1$ and $\bphi_2$, which is also consistent with DMDc.  
When $\bB$ is unknown, cDMDc and csDMDc both accurately identify $\bB$, regardless of the subspace it belongs to, as long as the columns of $\bB$ are sparse.  The algorithms do not accurately capture the $\bB$ matrix when it is not sparse (i.e., random actuation vector); however, in this case, DMDc also misidentifies the actuation vector.  

%\begin{figure*}[t]
%\begin{center}
%\begin{overpic}[trim = {0 0 0 5}, clip, width=.775\textwidth]{figures/Vorticity_field1}
%\put(57,23){Measurement window}
%\linethickness{2pt}
%\put(52,5){\line(1,0){38.5}}
%\put(90.5,5){\line(0,1){16}}
%\put(90.5,21){\line(-1,0){38.5}}
%\put(52,21){\line(0,-1){16}}
%\put(14,18){$t=0$}
%\put(3,20){(a)}
%\end{overpic}\\
%\begin{overpic}[trim = {0 0 0 5}, clip, width=.775\textwidth]{figures/Vorticity_field2}
%\linethickness{2pt}
%\put(52,5){\line(1,0){38.5}}
%\put(90.5,5){\line(0,1){16}}
%\put(90.5,21){\line(-1,0){38.5}}
%\put(52,21){\line(0,-1){16}}
%\put(14,18){$t=11.5$}
%\put(3,20){(b)}
%\end{overpic}
%\begin{overpic}[trim = {0 0 0 5}, clip, width=.775\textwidth]{figures/Vorticity_field3}
%\linethickness{2pt}
%\put(52,5){\line(1,0){38.5}}
%\put(90.5,5){\line(0,1){16}}
%\put(90.5,21){\line(-1,0){38.5}}
%\put(52,21){\line(0,-1){16}}
%\put(14,18){$t=21.5$}
%\put(3,20){(c)}
%\end{overpic}\\[2.ex]
%\begin{overpic}[trim = {0 0 0 6.75}, clip,width=0.775\textwidth]{figures/impulse}
%\put(3,23){(d)}
%\put(48,-2){Time (s)}
%\put(6,10.5){\begin{sideways}{Input}\end{sideways}}
%\end{overpic}
%\caption{(a)-(c) Instantaneous vorticity fields at different times, $t=0, 11.5, 21.5$s with the measurement window; (d) time series of the actuation specifying the pitching of plate. Black circles depict the time instants of the shown vorticity plots in (a)-(c). The delayed angle of pitch shows the shedding vortices entered the window at about $t=4.6s$.}\label{VorField}
%\vspace{-.2in}
%\end{center}
%\end{figure*}

\begin{figure*}[t]
\vspace{-.1in}
\begin{center}
\includegraphics[width=\textwidth]{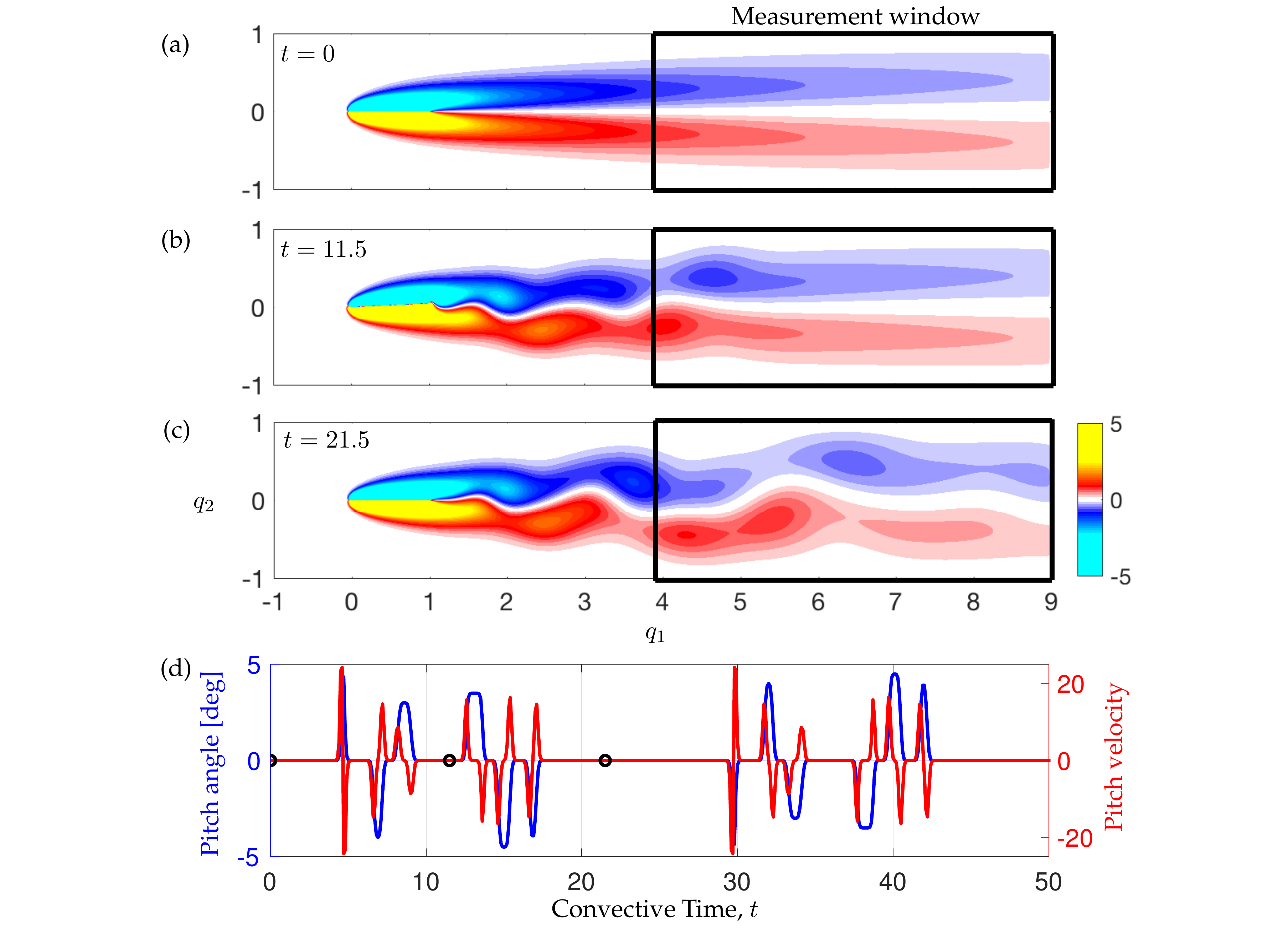}
		\caption{Simulation of a pitching airfoil: 
			(a)-(c) Instantaneous vorticity fields at different times $t=0, 11.5, 21.5$ and showing the measurement window, 
			(d) time series of the actuation inputs specifying the pitching of the plate. 
			Black filled circles depict the time instants of the shown vorticity plots in (a)-(c). }\label{VorField}
		\vspace{-.25in}
	\end{center}
\end{figure*}

\subsection{Fluid flow past a pitching plate}
In the second example, we apply the proposed compressive DMD with control algorithm to model the vorticity field downstream of a pitching plate at Reynolds number $Re=100$. 
This pitching airfoil has been studied previously in the context of reduced-order models for flow control~\cite{taira:07ibfs,taira:fastIBPM,Brunton2013jfm,Brunton2014jfs,Hemati2016aiaa}. 
The flow is simulated using the immersed boundary projection method (IBPM)\footnote{Code available at https://github.com/cwrowley/ibpm} method~\cite{taira:07ibfs,taira:fastIBPM} with a grid resolution of $799\times 159$ on a domain of size $10\times 2$, nondimensionalized by the plate length $L$. 
The flow is simulated with a time-step of $\Delta t=0.01$ dimensionless convective time units, nondimensionalized by the length $L$ and free-stream velocity $U$.  $251$ snapshots are sampled at a rate of $\Delta t=0.1$.  
In this example, the airfoil is rapidly pitched up and down between $\pm 5^\circ$ at irregular intervals in time, using the canonical pitch maneuver described in~\cite{canonical:2010}.  
Six rapid pitch maneuvers are performed, and then the data is symmetrized by concatenating a mirror image of these $251$ snapshots with the opposite signed vorticity, in an attempt to identify unbiased modes.   
Figure~\ref{VorField} illustrates the vorticity field at different times, $t = 0, 11.5, 21.5$. We focus on a downstream measurement window of the size $399\times141$ to mimic a PIV window. The actuation input takes $4.6$ convective time units to reach the observation window, and the actuation input, that used as input for cDMDc and DMDc, is shifted accordingly.  
Similar small amplitude pitching motions have been shown to be well-approximated with linear models~\cite{Brunton2013jfm}. 

\begin{figure*}[h]
\begin{center}
\includegraphics[width=\textwidth]{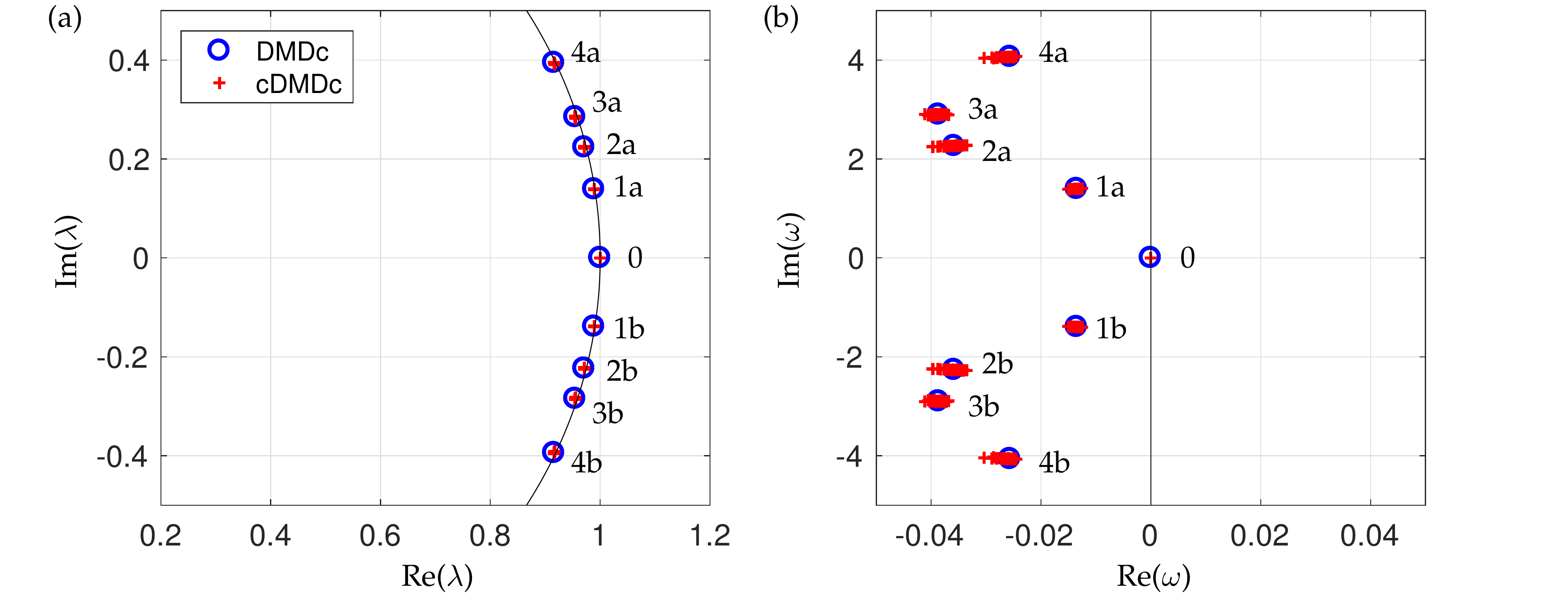}
\caption{Comparison of the spectrum in 
	(a) discrete time and 
	(b) continuous time, both obtained using DMDc and cDMDc with $10\%$ Gaussian random measurements. 
	The first $r=9$ modes are shown, ordered by their real part.}\label{Spectrum}
\end{center}
\end{figure*}

%The vortices shedding from the pitching plate upstream consists multiple-scale structures that oscillate at different frequencies. This example serves as a linear model that cDMDc will suitably decompose the modes form a subsampled measurements based on their decaying level, considering the impulse at the entrance of the subdomain. 
The eigenvalues of $\bAtilde$ and $\bAtildeY$ given by DMDc and cDMDc are shown in Fig.~\ref{Spectrum}.  
The cloud of cDMDc eigenvalues is generated from an ensemble of $50$ realizations using
$10\%$ compressed measurements (i.e., $p=0.1n$) based on Gaussian random projections. 
%The mode $0$ is stable across all snapshots and the others modes are decaying over time. 
{We use $r=9$ in this study to identify the spatiotemporal coherent DMD modes, and they capture $96\%$ of the total energy, based on the singular values. }
These nine modes are ordered by their decay rate (i.e., the real part) and the higher-order modes have larger frequencies of oscillation. 
The discrete-time eigenvalues are located close to the unit circle, and they are slightly stable, as seen in the continuous-time plot.  
The DMDc eigenvalues appear to have a decay rate that is too small, although the dynamics of transients are captured quite well.   
Nevertheless, the cDMDc eigenvalues agree well with the DMDc eigenvalues, which is the goal.
%, although there is slight variance in the decay rate.  

\begin{figure*}[t]
\centering
\includegraphics[width=\textwidth]{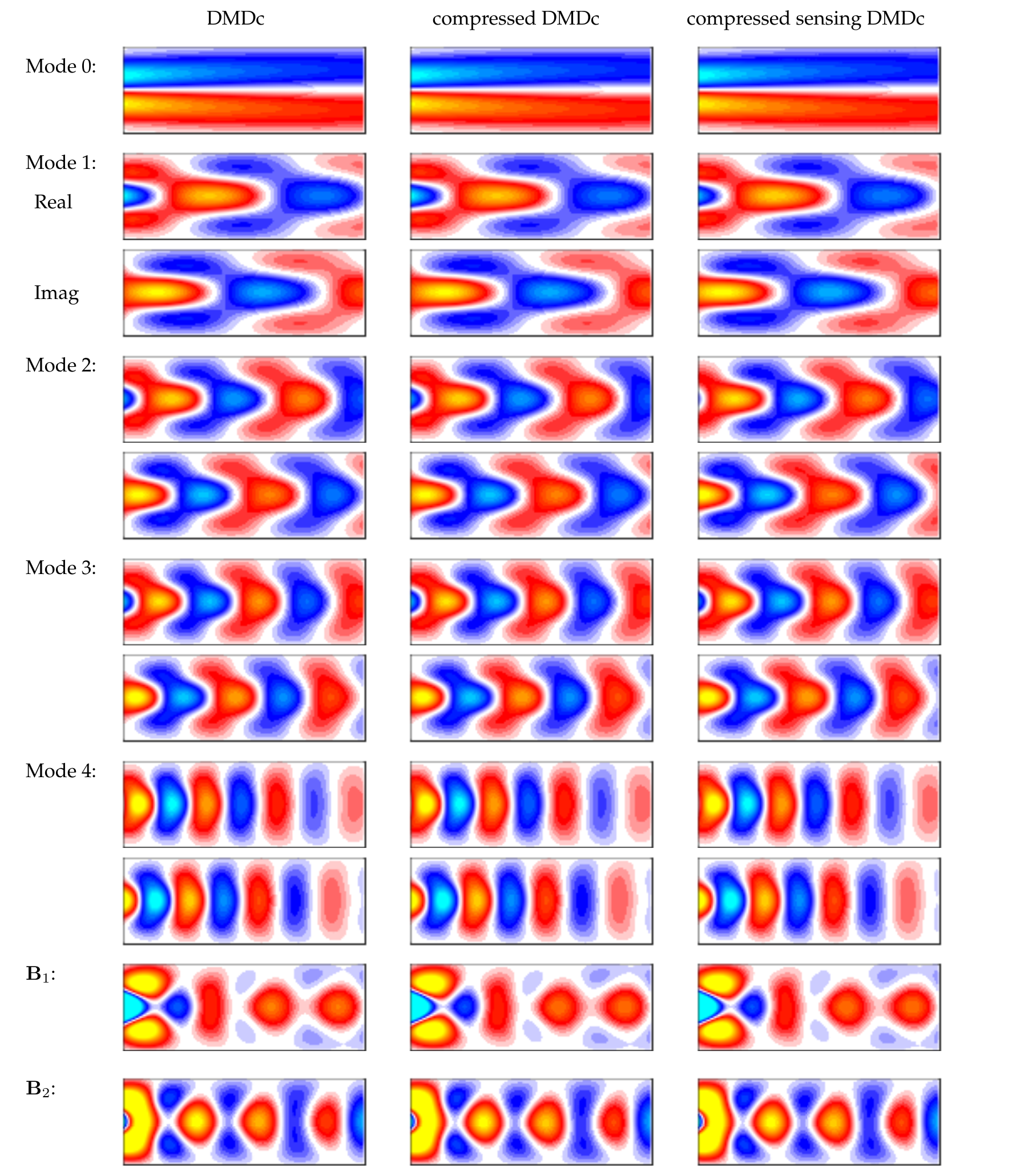}
\caption{Estimation of mode pairs (real and imaginary parts) and $\bB$ using DMDc, compressed DMDc and compressed sensing DMDc. 
	The cDMDc modes and actuation matrix $\bB$ (two columns, $\bB_1$ and $\bB_2$) are constructed using $p$ Gaussian random projections, where $p/n=0.1$ (i.e., 10\% compression ratio). 
	Note that the results of single pixel measurements are very similar (not shown).}\label{ModePairs}
\end{figure*}

\begin{figure*}
	\centering
\includegraphics[width=.95\textwidth]{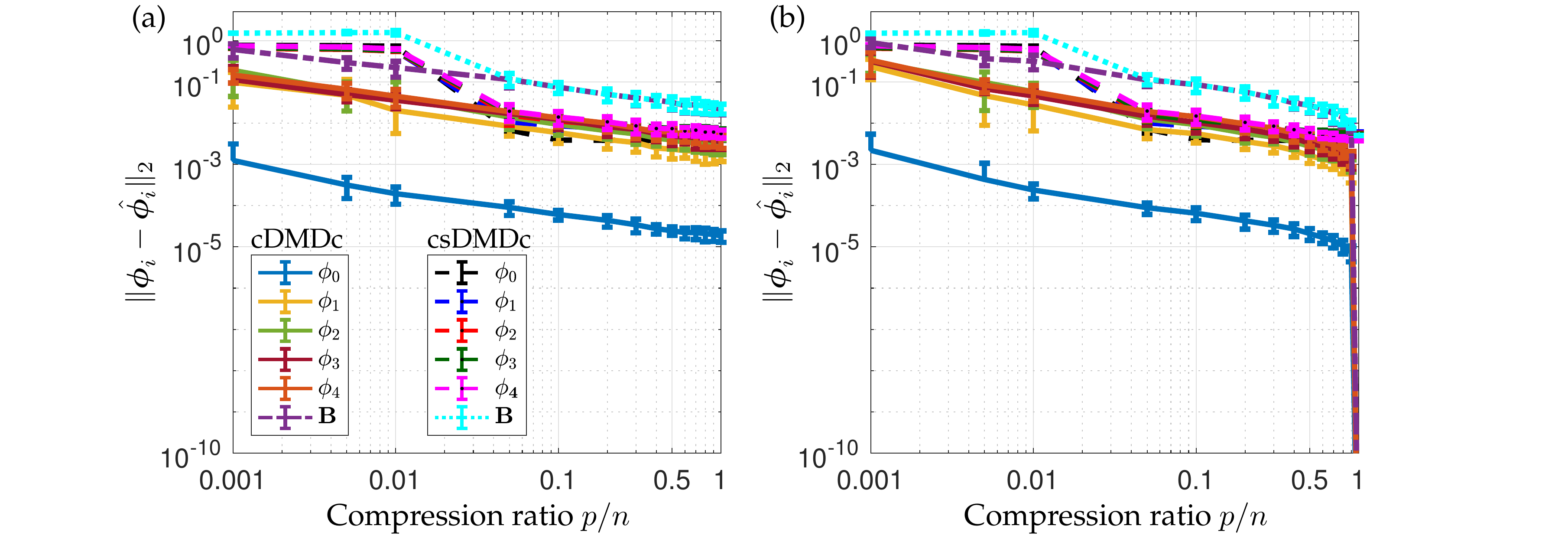}
	\caption{Error of estimated modes $\hat{\bphi}_i$ and actuation matrix $\hat{\bB}$ for increasing compression ratio using compressed DMDc and compressed sensing DMDc for 
		(a) Gaussian random projections and 
		(b) single pixel measurements. 
		The DMDc modes and actuation matrix are used as the reference, denoted by $\bphi$ and $\bB$, respectively.}\label{ErrorModes}	
	\centering
\includegraphics[width=.95\textwidth]{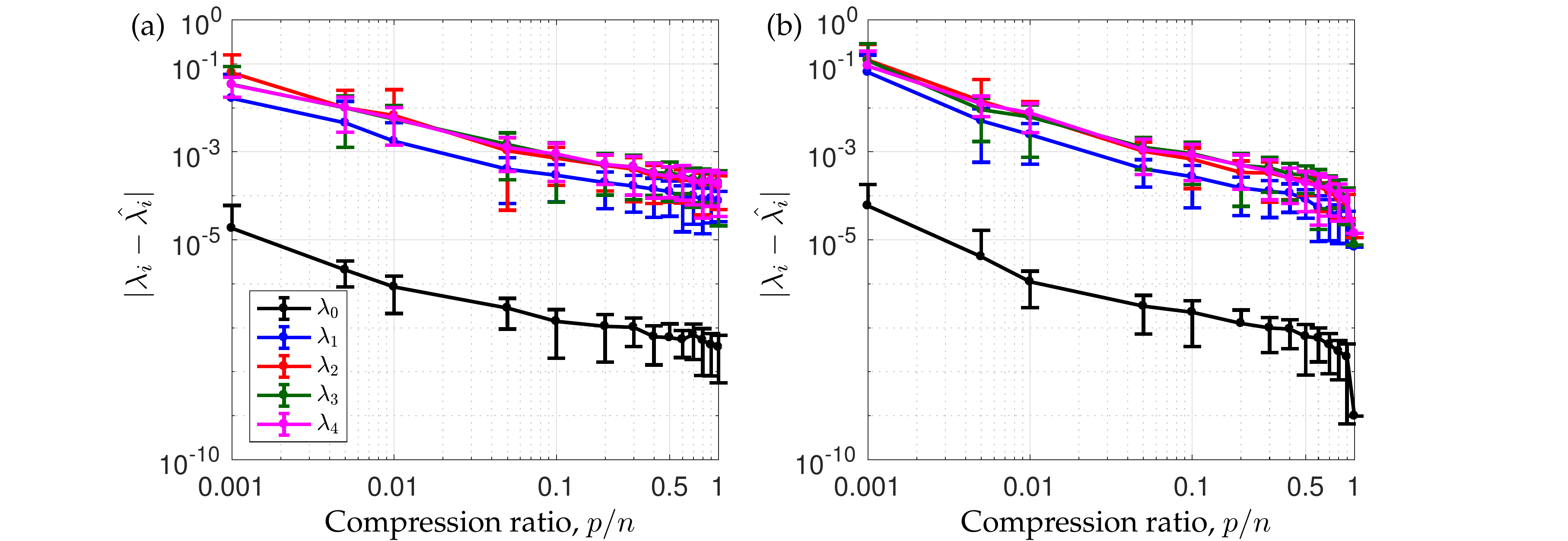}
	\caption{
		Error of eigenvalues for increasing compression ratio using 
		(a) Gaussian random projections, and 
		(b) single pixel measurements.  
		The DMDc eigenvalues are used as the reference $\lambda_i$.}\label{ErrorEigenvalues}
\end{figure*}

In Fig.~\ref{ModePairs}, we present the spatial structure of the DMDc mode pairs and the $\bB$ matrix. 
Similar DMD modes have been observed in the flow past a cylinder~\cite{Chen2012jns}. 
As the temporal frequency associated with an eigenvalue increases, the modes are characterized by higher spatial wavenumbers, which is characteristic of bluff-body flows.  
%We observe a slightly phase shift in Mode pairs $1-4$ and that may cause the slow convergence in the error plot of Figure~\ref{ErrorModes}. 
The accuracy of mode reconstruction is similar for Gaussian random projections and single-pixel measurements, as indicated in Fig.~\ref{ErrorModes}.    
%Using Gaussian random projections results in a slightly better reconstruction of the actuation matrix $\bB$ than using $10\%$ single pixel measurements, as indicated from Fig.~\ref{ErrorModes}.  
However, single pixel measurements may be less expensive and more realistic in real-world applications, and sampling $10\%$ of the original measurements results in a reconstruction accuracy of $90\%$ in this example.   

Taking the DMDc results as the reference, Fig.~\ref{ErrorEigenvalues} shows the error of the eigenvalues with increasing number of measurements used in cDMDc. Note that the eigenvalues are the same for compressed DMDc and compressed sensing DMDc, as shown in Algorithm~\ref{alg cDMDc} . 
The eigenvalue $\blambda_0$ corresponding to zero frequency is estimated with the greatest accuracy from the fewest measurements, presumably because this mode contains the most energy in the flow.  
The error in eigenvalues associated with other modes decreases logarithmically with increasing  compression ratio. 
Single pixel measurements have similar performance compared with Gaussian random measurements for small compression ratios. When $p=n$, the error goes to zero for single pixel measurement, since the measurement is an invertible permutation of the identity matrix, and cDMDc is equivalent to DMDc in this case.

{Successful reconstruction using compressed sensing relies on the sparsity of the  state in some transform basis.
Indeed, both $\bPhi$ and $\bB$ are sparse in the DCT basis.}
%Before performing compressed sensing on $\bPhiY$ and $\bBY$, we checked that both $\bPhi$ and $\bB$ are, in fact, sparse in the DCT basis. 
In general, the DMDc modes are more sparse than the actuation matrix, and we choose $K=300$ to ensure a good reconstruction of the modes and $\bB$ achieving an error of about $1\%$.  
In particular, we use the CoSaMP algorithm~\cite{Needell2010acm} to perform the $l_1$-minimization 
with $10$ iterations and the desired sparsity of $K=300$. 
The $L^2$ errors between the compressive DMDc and DMDc modes are shown in Fig.~\ref{ErrorModes}. The convergence of error versus compression ratio in compressed sensing DMDc is much slower than that in compressed DMDc, especially for the zero frequency mode. 

The necessary number of measurements 
%can be approximated from 
is given from theory to be
$p\sim 4K\log_{10}(n/K) \approx 2728$, which corresponds to a compression ratio of $5\%$. 
This matches the observation that the error curves plateau %to a steady state 
at around $p/n = 0.05$. 
%A similar behavior at $20\%$ compression ratio is also observed as in Fig.~\ref{ErrorEigenvalues}. 
%To reconstruct $\bB$, Gaussian random projections are more effective than single pixel measurements.  
%For example, an accuracy above $90\%$ in the prediction of $\bB$ is achieved with $\sim 10\%$ Gaussian random measurements, in contrast to similar performance with over $\sim 50\%$ single pixel measurements.  
%\textcolor{red}{(The last sentence contradicts and earlier statement in the 3rd paragraph!!!)}
Overall, the error for compressed measurements using Gaussian random projections is comparable with using single pixel measurements.

%%%%%%%%%%%%%
%%%% DISCUSSION
%%%%%%%%%%%%%
%\input{Sec5}

%%%%%%%%%%%%
%%% CONCLUSION
%%%%%%%%%%%%
\section{Conclusions}\label{sec:conclusion}
\vspace{-.15in}
In summary, we have presented a unified framework for compressive system identification based on the dynamic mode decomposition.  
First, we describe the previously developed compressed sensing DMD (csDMD) and DMD with control (DMDc) algorithms in a common mathematical framework, providing algorithmic implementation details.
Next, we show how it is possible to construct reduced-order models from compressed measurements and then reconstruct full-state modes corresponding to the reduced states via compressed sensing. 
This lifting procedure adds interpretability to otherwise black-box models.  
The compressed DMD with control algorithm is demonstrated on two example systems, including a high-dimensional discretized simulation of fluid flow past a pitching airfoil.  
In both cases, accurate modal decompositions are achieved with surprisingly few measurements, showcasing the efficacy of the proposed method. 
In addition, we have released our entire code base to promote reproducible research and reduce the barrier to implement these methods.  

There are a number of important extensions and future directions that arise out of this work. 
First, it will be interesting to further investigate the relationship between the controllable and observable subspaces and full-state recovery via compressed sensing.  
The goal is a generalized theory that combines the notion of controllability and observability, based on the structure of the $\bA$, $\bB$, and $\bC$ matrices, and the notion of sparse signal recovery, via the structure of the low-rank embedding $\bP$.  
It may also be possible to extend results to nonlinear estimation~\cite{Surana2016cdc,Surana2016linear} and control~\cite{Proctor2016KIC,Korda2016MPC,Kaiser2017arxiv} through the connection to the Koopman operator.  
In addition, the methods described here are directly applicable to experimental measurements~\cite{Sharma2016prf}, as they do not require access to a model of the system. 
It may be possible to significantly reduce the required spatial resolution in experiments, improving the effective bandwidth and enabling the characterization of faster flow phenomena.  
A sparsifying POD basis may be obtained first with non-time-resolved measurements at full resolution. 
It is then possible to collect many fewer spatial measurements at much higher temporal resolution, identify a reduced-order model, and characterize the full-state modes in the offline library.  
This also suggests that it may be possible to combine space and time compressed sensing strategies for DMD.  

The growing intersection of dynamical systems, machine learning, and advanced optimization are driving tremendous innovations in the characterization and control of complex systems~\cite{Duriez2016book,Kutz2016book}.  
Although data is becoming increasingly abundant, there remain applications such as feedback flow control, where real-time measurements are costly and control decisions must be made with low latency to ensure robust performance.  
In these applications, techniques that strategically select sensor data into the most relevant information will enable higher performance in more sophisticated flow control applications. 
It is likely that the methods developed here may be combined with principled sensor selection methods to promote enhanced sparsity based on learned structures and patterns~\cite{Brunton2016siap,Bai2015springer,Kaiser2016arxiv,Manohar2017csm}.  
Moreover, it may also be possible to enforce known physics or symmetries in the regression procedure, as in~\cite{Loiseau2016arxiv}, to improve model performance and accelerate learning from data.

\section*{Acknowledgements}
SLB acknowledges support from the Air Force Office of Scientific Research (AFOSR FA9550-16-1-0650; Program Manager: Dr. Douglas R. Smith) and the Army Research Office (ARO W911NF-17-1-0118; Program Manager: Dr. Matthew Munson).  
We would also like to thank several people for valuable discussions about sparse sensing, dynamic mode decomposition, and modeling high-dimensional fluids, including Bing Brunton, Krithika Manohar, Igor Mezi\'c, Bernd Noack, and Sam Taira.

%%%%%%%%%%%%
%%% BIBLIOGRAPHY
%%%%%%%%%%%%
%\bibliographystyle{model1-num-names} 
%\bibliography{Literature.bib}
%\bibliographystyle{myplain}
%\bibliography{main}
%\renewbibmacro{in:}{}
%\renewcommand{\bibfont}{\footnotesize}
%%\begin{spacing}{.95}
%%\setstrech{0.8}
%\setlength\bibitemsep{.8pt}
%\printbibliography
%%\end{spacing}
\bibliographystyle{siam}
\bibliography{Literature}

%\end{multicols}
\end{document}